\documentclass{emulateapj}
\usepackage{amsbsy}
\usepackage{graphicx} 

\def\bs{\boldsymbol}
\def\gsim{\;\rlap{\lower 2.5pt
\hbox{$\sim$}}\raise 1.5pt\hbox{$>$}\;}
\def\lsim{\;\rlap{\lower 2.5pt
\hbox{$\sim$}}\raise 1.5pt\hbox{$<$}\;}

\lefthead{Pan \& Scannapieco}  \righthead{Mixing in Supersonic Turbulence}

\begin{document}
\title{Mixing in Supersonic Turbulence}
\author{Liubin Pan \& Evan Scannapieco}
\affil{School of Earth and Space Exploration,  Arizona
  State University, P.O.  Box 871404, Tempe, AZ, 85287-1404.}
  
\begin{abstract}

In many astrophysical environments,  mixing of heavy 
elements occurs in the presence of a supersonic turbulent 
velocity field. Here we carry out the first systematic 
numerical study of such passive scalar mixing in 
isothermal supersonic turbulence.  Our simulations show that 
the ratio of the scalar mixing timescale, $\tau_{\rm c}$,
to the flow dynamical time, $\tau_{\rm dyn}$ (defined as the 
flow driving scale divided by the rms velocity),  
increases with the Mach number, $M$, for $M \lsim3$, 
and becomes essentially constant for $M \gsim3.$  
This trend suggests that compressible modes are less 
efficient in enhancing mixing than solenoidal modes. However, 
since the majority of kinetic energy is contained in solenoidal 
modes at all Mach numbers, the overall change in $\tau_{\rm c}/\tau_{\rm dyn}$ is 
less than 20\% over the range $1 \lsim M \lsim 6$. At all Mach numbers, if pollutants are 
injected at around the flow driving scale, $\tau_{\rm c}$ is close to 
$\tau_{\rm dyn}.$ This suggests that scalar mixing is 
driven by a cascade process similar to that of the velocity field. 
The dependence of $\tau_{\rm c}$ on the length scale at 
which pollutants are injected into flow is also consistent 
with this cascade picture. Similar behavior is found for 
the variance decay timescales for scalars without continuing 
sources. Extension of the scalar cascade picture to the supersonic 
regime predicts a relation between the scaling exponents of the 
velocity and the scalar structure functions, with the scalar structure 
function becoming flatter as the velocity scaling steepens with 
Mach number. Our measurements of the volume-weighted velocity 
and scalar structure functions confirm this relation for $M\lsim 2,$ 
but show discrepancies at $M \gsim 3$,  which arise 
probably because strong expansions and compressions 
tend to make scalar structure functions steeper.  

\end{abstract}

\keywords{}

\section{Introduction}

Understanding mixing in turbulent flows is essential 
to interpreting a wide range of observations
including:  the metallicity dispersion in open 
clusters (e.g., Friel \& Boesgaard 1992; Quillen 2002; DeSilva et al.\ 2006),
the cluster to cluster metallicity scatter (Twarog et al.\ 1997),  
the abundance scatter along different 
lines of sight in the interstellar medium (ISM) (Meyer et al.\ 1998, Cartledge et al.\ 2006), 
the scatter in metallicities and abundance ratios in coeval 
field stars (e.g., Edvardsson et al.\ 1993; Nordstrom et al.\ 2004; 
Reddy et al.\ 2003, 2006).
The degree of chemical inhomogeneity in these objects is  
controlled by the efficiency at which interstellar 
turbulence transports and mixes metals from 
supernova explosions and stellar winds (Scalo \& Elmgreen 2004).  
This mixing process is also essential to the 
pollution of primordial gas by the first generation 
of stars in early galaxies (e.g., Pan \& Scalo 2007), 
and the transition from Population III to Population 
II star formation (e.g., Scannapieco et al.\ 2003).

Mixing in incompressible turbulence 
has been extensively studied in the fluid 
dynamics literature, where it is usually referred 
to as ``passive scalar turbulence" 
(e.g., Shraiman \& Siggia 2000). 
The classic picture, known as the Obukohov-Corrsin 
phenomenology, is a cascade of scalar fluctuations, 
which is caused by the stretching of the concentration field 
by the velocity field. This produces structures 
of progressively smaller size, down to the 
diffusion scale at which molecular diffusion 
homogenizes and erases the fluctuations. In fact,  
molecular diffusion is the only process responsible 
for true mixing, but it operates very slowly at the 
large scales at which the scalar sources are injected. 
By providing a path from the scalar injection scale 
to the diffusion scale, the cascade greatly 
accelerates the mixing process. 

Turbulent stretching is faster in smaller eddies, 
and the mixing timescale is essentially given 
by the timescale of eddies at the scalar injection 
scale, which is independent of the molecular 
diffusion coefficient. Scalar structures at small 
scales are determined by the velocity structures 
that control the scalar cascade. For an incompressible 
flow with a Kolmogorov spectrum, the Obukohov-Corrsin 
phenomenology predicts a $-5/3$ scalar spectrum or 
a 2/3 power law for the 2nd order scalar structure 
function in the ``inertial-convective" scale range between 
the driving scales and the dissipation scales of the 
velocity and scalar fields. 

Unlike most terrestrial flows, turbulence in the ISM is 
usually supersonic, due to efficient radiative cooling.  
Therefore, understanding passive scalar physics in supersonic 
turbulence is essential for studying mixing in the 
interstellar medium. In contrast to the extensive efforts 
on mixing in incompressible turbulence, 
mixing in supersonic flows has received little 
investigation. To our knowledge, there have 
been only two numerical studies of transport and mixing of 
chemical elements in supersonic interstellar turbulence, 
and these have not addressed the underlying 
physics of turbulent mixing. First, de Avillez and Mac 
Low (2002) used simulations to compute the timescale 
for the variance decay of concentration fields with no continuing 
sources in supernova-driven interstellar turbulence. 
However, they did not give a detailed study or discussion of 
the central  physical issues in turbulent mixing, such as the 
production of small-scale scalar fluctuations and their 
effect on the mixing timescales. Second, Klessen and Lin (2003) 
considered the dispersal of tracer particles in supersonic 
turbulence, but did not study mixing in the sense of 
homogenization of concentration fluctuations.    

The goal of the present work is to undertake the first 
systematic study of the physics of mixing in 
supersonic turbulence. To approach this question, 
we have carried out numerical simulations for 
turbulent flows at different Mach numbers and 
measured mixing timescales, structure functions, 
and other statistical quantities, drawing detailed conclusions 
that can be applied in a wide range of astrophysical contexts. 

We are particularly interested in two key issues. 
First, we assess whether the cascade picture 
can be successfully applied to passive 
scalars in supersonic flows. 
Extending the Obukhov-Corrsin theory to mixing 
in supersonic turbulence predicts a 
relation between the scalar scaling and 
the velocity scaling, corresponding to the 
assumption that the cascade timescale 
is proportional to the eddy turnover time at each scale. 
With this relation, the scalar scaling at a given 
Mach number would be fixed 
by the velocity scaling (see \S 2.2). In particular, 
the scalar scaling would be predicted to flatten with 
increasing Mach numbers as the velocity scaling 
becomes steeper.  Also the cascade picture 
predicts the mixing timescale would be proportional 
to the turnover time of eddies at the scalar source 
scale. The applicability of the cascade picture 
will be examined by comparing these 
predictions against our numerical 
simulation results. 

The second issue is the role of the compressible 
modes, which contain a significant fraction of  
kinetic energy in highly supersonic turbulence.  
Unlike solenoidal modes which change 
the scalar geometry by stretching,  
compressible modes squeeze and 
expand the scalar field, which may have important 
effects on the scalar spectrum and structure function. 
Of special interest is the efficiency at which 
compressible modes produce small-scale scalar structures, 
which determines the contribution of these modes to the 
mixing rate.  
The contributions to mixing by two types of modes 
may be significantly different,  and this would have 
consequences for the mixing timescale as a function 
of Mach number.  
The possibility that compressible modes play 
a different role in mixing has been suggested 
by Gawedzki \& Vergassola (2000), who predicted 
that the behavior for passive scalars in highly 
compressible flows may be completely different 
from that in incompressible turbulence.   
A comparison of mixing timescales in our 
simulated flows at different Mach numbers 
will provide clues to the role of compressible 
modes in  mixing in compressible turbulence.

The structure of this work is as follows. In \S 2,  
we review the physics of turbulent mixing, and 
point out the important issues that need to be 
addressed for a full understanding of mixing 
in supersonic turbulence. 
In \S 3, we describe the numerical method and 
the simulation setup used in our study, and results 
are presented in \S 4. We discuss the role of compressible 
modes in mixing in detail in \S 5, and our major 
results are summarized in the conclusion section (\S6).  
   
\section{Turbulent Mixing and Kinetic Energy Dissipation}

\subsection{Physics of Mixing}
We study mixing of pollutants in a turbulent flow $\bs{v}(\bs{x}, t)$ 
with a density field $\rho(\bs{x},t)$. The mixing process 
corresponds to the homogenization of the concentration field 
$C(\bs{x},t)$, defined as the ratio of  the pollutant density 
to the local flow density. 
The evolution of the  concentration field is described by the advection-diffusion equation, 
\begin{equation}
\partial_t  C +   v_i \partial_i C =  \frac{1}{\rho} \partial_i (\rho \kappa \partial_i C) +  S(\bs{x},t),
\label{eq:Ceq}
\end{equation}   
where $\kappa$ denotes the molecular diffusivity, and the 
term $S(\bs{x},t)$ represents the scalar sources.
To understand the physics of turbulent mixing, it is crucial 
to recognize the role of the turbulent velocity. Intuitively, 
a velocity field redistributes the concentration field 
and changes its spatial configuration, but does not affect 
the mass fraction at a given concentration level, since the 
tracer particles simply follow the flow elements. This is 
true for any velocity field no matter how complex. 
Therefore, the velocity field itself does not mix, because 
true mixing means spatial smearing of pollutants, 
which can be caused only by molecular diffusion.  

The observation that the velocity field is not responsible for 
true mixing can be formally confirmed from the equation for the variance of 
the concentration fluctuations, which is usually referred to as the 
scalar energy in the turbulence literature. To correctly reflect this 
intuition in compressible flows, one needs to consider the 
{\em density-weighted} variance, $\langle \tilde{\rho} C^2 \rangle,$ where 
the density-weighting  factor $\tilde{\rho}$ is the ratio of $\rho$ to the 
average flow density,  $\bar{\rho}$, in the flow, and $\langle \cdot \cdot \cdot \rangle$ 
denotes an ensemble average, which is equal to the average 
over the flow domain for statistically homogeneous flows, as 
in our simulations.

The equation for the density-weighted variance can be 
derived from eq.\ (\ref{eq:Ceq}) with the help of the continuity 
equation, yielding 
\begin{eqnarray}
\partial_t  \langle \tilde{\rho} C^2 \rangle + \partial_i \langle \tilde{\rho} C^2 v_i \rangle 
&=& 2 \langle \partial_i (\tilde{\rho} \kappa C \partial_i C) \rangle  \nonumber \\
& & -  2  \langle \tilde{\rho} \kappa (\partial_i C)^2 \rangle  
+  2 \langle \tilde{\rho} S C \rangle.  
\label{eq:Cvariance}
\end{eqnarray}
The second term on the left hand side of this equation is an 
advection term, which corresponds to the spatial transport of 
concentration fluctuations between different regions.  
In the statistically homogeneous case, this term vanishes, 
and without it, eq.\ (\ref{eq:Cvariance}) does not have an explicit 
dependence on the velocity field. This proves the intuition that 
the velocity field does not truly mix.  In fact, this is also true in  
an inhomogeneous flow. While the advection term is 
not zero in this case, it is a surface term, and thus its 
integral over the entire flow domain is zero. 
This means that the advection term conserves the global 
scalar variance\footnotemark\footnotetext
{Without density-weighting, there would be a term $-\langle  C^2 \partial_i v_i \rangle$ 
in the equation for the (volume-weighted) scalar variance. 
This term is not zero in general, and represents the change 
in the volume fraction of the flow elements with a given 
concentration due to compressions and expansions. 
This change in the volume fraction has nothing to do 
with real mixing, and it is more appropriate to use the 
density-weighted variance, which is conserved by the advecting velocity field.}.   
 
On the right hand side of eq.\ (\ref{eq:Cvariance}), the last term 
is the source term, which corresponds to the variance increase 
due to the injection of new pollutants. The other two terms on the 
right hand side of this equation arise from the diffusion term in eq.\ (\ref{eq:Ceq}). 
The first of these is again a surface term, which vanishes in the homogeneous case and 
conserves the global scale variance in the inhomogeneous case. 
The term $-2\langle \tilde{\rho} \kappa (\partial_i C)^2 \rangle$ 
represents scalar dissipation. This term is negative definite  
and thus monotonically reduces the scalar variance,  
corresponding to homogenization of the scalar 
fluctuations by molecular diffusion. We denote the 
dissipation rate as $\bar{\epsilon}_{\rm c} \equiv 2\langle \tilde{\rho} \kappa (\partial_i C)^2 \rangle$ 
and define a scalar dissipation timescale $\tau_{\rm c} \equiv \langle \tilde{\rho} C^2 \rangle/\bar{\epsilon}_{\rm c}$ 
to characterize it. The linear dependence on $\kappa$ 
in the definition of $\bar{\epsilon}_{\rm c}$ may leave
an impression that the scalar dissipation rate strongly 
depends on $\kappa$. However, that is not true 
because the scalar gradient in the definition also 
depends on $\kappa$. With decreasing $\kappa$, larger scalar gradients can develop, 
which may compensate the decrease in 
$\kappa$. As seen from the physical picture 
below and in \S 2.2, the dissipation rate and the mixing 
timescale are essentially independent of the molecular diffusivity 
\footnotemark\footnotetext{The mixing timescale 
may have a weak dependence on $\kappa$ if 
the Schmidt number (see definition below) is much larger than 1.}. 
 
The molecular diffusivity $\kappa$ is tiny in most 
practical environments, thus the dissipation rate is 
significant only if the scalar gradient is large, meaning 
that true mixing occurs only at very small scales.  
As mentioned earlier, at the injection scale of scalar sources, 
the molecular diffusion is usually very slow, and a significant 
mixing rate needs a velocity field to produce small-scale scalar 
structures, or equivalently,  large concentration gradients. 
In an incompressible turbulent flow, the velocity field 
stretches, contracts and folds the scalar field, leading 
to scalar structures at smaller and smaller scales.   
The size of the scalar structures keeps decreasing toward 
the diffusion scale where the molecular diffusion efficiently 
homogenizes the structures. Since $\kappa$ is small, a 
wide scale separation usually exists between the 
scalar source size and the diffusion scale.  

The diffusion scale is essentially the scale at which  
the diffusion term in eq.\ (\ref{eq:Ceq}) becomes 
larger than the advection term, and thus can derived 
by comparing these two terms accounting for their 
scale dependences. The estimate of the diffusion scale 
depends on the Schmidt number, $Sc$, defined 
as the ratio of viscosity to diffusivity, which determines the 
diffusion scale relative to the viscous dissipation 
scale of the velocity field (the scale at which  
kinetic energy is removed by viscosity). The 
derivation of the diffusion scale at different $Sc$ values 
can be found in Monin and Yaglom (1975) and 
Lesieur (1997). For heavy elements 
in a neutral interstellar medium, the diffusivity 
is expected to be smaller than the viscosity 
due to the larger cross section of heavy 
elements. However, for relatively light atoms like 
oxygen, the diffusivity is probably of the same order 
as the viscosity, and the Schmidt number is close to unity.    

A second important dimensionless number for 
passive scalars is the Peclet number, $Pe$. It is 
defined as $UL/\kappa$ where $U$ and $L$ are 
characteristic velocity and length scale of the 
turbulent flow. The Peclet number is related to the 
Reynolds number, $Re$, by $Pe= Sc Re$. Therefore, 
for diffusing species with $Sc \approx 1$,  
the Peclet number  is close to the 
Reynolds number, $\gsim 10^6$, in typical interstellar 
clouds.
 
In summary,  a turbulent velocity field acts as a 
catalyst that enhances the mixing efficiency by 
feeding molecular diffusion with large gradient 
structures. The mixing timescale, $\tau_{\rm c}$, 
is thus determined by the rate at which 
the turbulent velocity field brings the scalar 
structures down to the diffusion scale.  
It is essentially independent of $\kappa$ for the 
case with $Sc \lsim 1$.  

\subsection{Scalar Cascade}

The classic picture for the generation of 
scalar structures at small scales in 
incompressible turbulence is a cascade similar to the energy 
cascade of the velocity field (e.g., Lesieur 1997). 
A constant flux of the scalar energy occurs 
along the cascade over the convective range,
defined as the range of scales between the 
characteristic length scale of the scalar 
sources and the diffusion scale. In this range, the scalar 
fluctuations are regulated primarily by the 
advecting velocity.  The flux feeds the dissipation 
at the diffusion scale and is thus expected to 
be equal to the average scalar dissipation rate. 
Assuming the timescale for a cascade step is of the order of the 
eddy turnover time at a given scale leads 
to the following scaling for the concentration difference, 
$\delta C(l)$, over a separation $l$, 
\begin{equation} 
\delta C(l)^2 \simeq \bar{\epsilon_{\rm c}} \frac{l}{\delta v(l)}, 
\label{eq:scaling}
\end{equation}
where ${\delta v(l)}$ is the velocity difference over 
$l$, and $\bar{\epsilon}_{\rm c}$ is the scalar dissipation rate.

In incompressible turbulent flows, we have the Kolmogorov 
scaling for the velocity difference, $\delta v(l)$, in the inertial 
range, i.e., $\delta v(l) \simeq \bar{\epsilon_{\rm v}}^{1/3}l^{1/3}$, 
where $\bar{\epsilon_{\rm v}}$ is the average dissipation rate of 
kinetic energy. With this velocity scaling, eq.\ (\ref{eq:scaling}) 
gives $\delta C(l)^2 \simeq  \bar{\epsilon}_{\rm v}^{-1/3} \bar{\epsilon}_{\rm c} l^{2/3}$ 
in the inertial-convective range (i.e., the intersection of the 
inertial range for the velocity field and the convective range 
for the scalar field). 
This scaling for the scalar difference, usually referred to as the Obukhov-Corrsin scaling law, 
corresponds to a $k^{-5/3}$ scalar spectrum. 
This spectrum has been confirmed for passive scalars in  
isotropic incompressible turbulence by both 
experiments and numerical simulations (Monin and Yaglom 1975; Sreenivasan 1996;  Mydlarskiand 
and Warhaft 1998;  Yeung et al.\ 2002; Watanabe \& Gotoh 2004). 

On the other hand,  the scalar structure function and spectrum 
have not been studied in supersonic turbulence. Therefore it is 
unknown whether eq.\ (\ref{eq:scaling}) originally 
proposed for passive scalar mixing in 
incompressible turbulence is valid in supersonic turbulence. 
The scaling of the velocity difference in supersonic 
turbulence has been shown to be significantly steeper than 
the Kolmogorov scaling (e.g., Kritsuk et al. 2007). Thus, 
if eq.\ (\ref{eq:scaling}) holds for turbulent flows 
at any Mach number, one would expect the scalar structure 
functions to become flatter with increasing Mach number.   
We will measure the scaling exponents, $\zeta_{\rm v}$ and $\zeta_{\rm c}$, of the 2nd  
order velocity structure functions ($\langle \delta v(l)^2 \rangle \propto l^{\zeta_{\rm v}} $) 
and the scalar structure functions ($\langle \delta C(l)^2 \rangle \propto l^{\zeta_{\rm c}} $)  
from our simulations at different Mach numbers.  
Eq.\ (\ref{eq:scaling}) predicts the following relation between the 
two exponents,  
\begin{equation} 
\zeta_{\rm c}  \simeq 1- \zeta_{\rm v}/2,
\label{eq:exponents}
\end{equation}
which will be tested against our simulation data.  

Eq. (\ref{eq:scaling}) also suggests that, if the 
scalar field and the flow are driven at similar scales, 
the dissipation timescales for scalar fluctuations 
and for kinetic energy will be similar. They are both 
determined by the turnover time of largest eddies 
at the driving/source scale, which is of the order of the 
dynamical timescale, because the cascade becomes 
progressively faster at smaller scales, and the
most time is spent at large scales. Numerical 
simulations confirmed that these two timescales are similar in 
incompressible turbulence (see, e.g.\
Donzis,  Sreenivasan, \& Yeung et al.\  2005 and references therein)
and that the scalar dissipation timescale is about half the energy 
dissipation timescale. Furthermore, in the supersonic case, the 
energy dissipation rate has been found to be similar to that 
in incompressible turbulence, i.e., close to the dynamical 
timescale (Stone et al.\ 1998; Mac Low 1998; Padoan and Norlund 1999), 
implying an energy cascade 
similar to that in incompressible turbulence.  
However, in the supersonic case, the dissipation timescale  
for forced scalars has never been computed.

\subsection{Kinetic Energy Dissipation}

Our simulations also give us an opportunity to explore the dissipation 
of kinetic energy in supersonic turbulence. The momentum 
equation in a driven turbulent flow is given by,     
\begin{equation}
\partial_t  v_i +   v_j \partial_j v_i = - \frac{\partial_i p}{\rho} + \frac{1}{\rho} \partial_j (\sigma_{ij}) +  
f_i (\bs{x}, t), 
\label{eq:velocity}
\end{equation} 
where  $p(\bs{x}, t)$ is the pressure, $\bs{f}(\bs{x}, t)$ is the driving force, and $\sigma_{ij}$ is 
the viscous stress tensor.  For an ideal gas, the bulk viscosity is 
negligible and the viscous tensor $\sigma_{ij} =  \frac{1}{2} \rho \nu ( \partial_i v_j +\partial_j v_i - \frac{2}{3} \partial_k v_k \delta_{ij}) $
where $\nu$ is the kinematic viscosity.  
 
From the momentum equation, eq.\ (\ref{eq:velocity}), and the 
continuity equation, we can derive an equation for the average kinetic energy per unit 
mass, $\langle \frac{1}{2} \tilde{\rho} v^2 \rangle$,  
\begin{eqnarray}
\partial_t  \langle  \frac{1}{2} \tilde{\rho} v^2 \rangle  
& =&  \frac{1}{\bar{\rho}} \left \langle p \partial_i v_i \right \rangle    +  \langle \tilde{\rho} f_i v_i \rangle \nonumber \\
& & - \frac{1}{2} \left \langle  \tilde{\rho} \nu \left( \partial_i v_j +\partial_j v_i - 
\frac{2}{3} \partial_k v_k \delta_{ij} \right)^2   \right \rangle
\label{eq:keperunitmass}
\end{eqnarray}
The three terms on the right hand side of eq.\ (\ref{eq:keperunitmass}) represents 
the pdV work,the energy injection from the driving force, and  the viscous dissipation of kinetic energy, 
 respectively.   

The pdV term, which vanishes in incompressible flows, 
is a two-way energy exchange between kinetic energy 
and thermal energy. Although the conversion 
by pdV work between the two energy forms 
is reversible, the exchange rates in the two directions 
can be asymmetric. In fact, the term preferentially 
converts kinetic energy to heat due to an anticorrelation 
between the density (or pressure) and the velocity 
divergence. Consider, for example, an expanding 
region and a converging region with the same $|\nabla \cdot \bs{v}|$. 
In the converging region, the density is 
typically larger, and thus the rate at which the 
pdV work converts kinetic energy to heat 
is faster than the conversion from 
thermal to kinetic energy in the corresponding 
expanding region. Therefore, in supersonic turbulence,  
the pdV work provides another mechanism 
for kinetic energy loss in addition to 
viscous dissipation. We denote the rate of kinetic 
energy loss by this mechanism as 
 $\bar{\epsilon}_{\rm p} \equiv - \langle \frac{p}{\bar{\rho}} \partial_i v_i \rangle$,  
and point out that this effect has not been explicitly 
considered in previous studies.  

Viscous dissipation, corresponding to the second term 
on the right hand side of eq.\ (\ref{eq:keperunitmass}),
is an irreversible conversion of kinetic energy to heat.  
If $\rho \nu$ is constant, which is true if the flow temperature 
is roughly constant, the overall average kinetic energy dissipation
rate in the flow can be written as $\bar{\epsilon}_{\nu} =  \tilde{\rho}\nu [\langle  (\nabla \times \bs{v})^2  \rangle  +  \langle \frac{4}{3} 
(\nabla \cdot \bs{v} )^2 \rangle]$ using integration by parts. 
We define a viscous dissipation timescale, $\tau_{\rm diss}$, 
as $\frac{1}{2} \langle \tilde{\rho} v^2 \rangle/\bar{\epsilon}_{\nu},$ 
and point out that this timescale for kinetic energy loss purely by 
viscous dissipation, to our knowledge, has also never been evaluated in 
supersonic turbulence.

In fact, the only timescale for kinetic energy loss estimated in earlier 
studies is the overall dissipation timescale $\tau_{\rm v}$ that characterizes 
the total rate for the kinetic energy loss by both pdV work and viscous 
dissipation, $\tau_{\rm v} = \frac{1}{2} \langle \tilde{\rho} v^2 \rangle/(\bar{\epsilon}_{\nu} + \bar{\epsilon}_{\rm p})$
(Stone et al.\ 1998; Mac Low 1998; Padoan \& Norlund 1999; Lemaster \& Stone 2009).
For example,  the timescale for  kinetic energy loss, $t_{\rm diss}$ defined in Lemaster \& Stone (2009) 
corresponds to the timescale $\tau_{\rm v}$ defined here.  Below 
we will evaluate both $\tau_{\rm diss}$ and $\tau_{\rm v}$, separating out 
the relative importance of kinetic energy loss by pdV work and viscous 
dissipation as a function of Mach number.

\section{Methods}

To study mixing in supersonic turbulence,
we simulated hydrodynamic turbulent flows at various Mach 
numbers using the FLASH code (version 3.2), a multidimensional 
hydrodynamic code  (Fryxell et al.\  2000) that solves the Riemann problem on a
Cartesian grid using a directionally-split  Piecewise-Parabolic Method
(PPM) solver (Colella \& Woodward 1984; Colella \& Glaz 1985; Fryxell,
M\" uller, \& Arnett 1989). The hydrodynamic equations and the advection-diffusion 
equation were solved on a fixed 512$^3$ grid for a domain of unit size with 
periodic boundary conditions.  We adopted an isothermal equation of state  in all our 
simulated flows, which was obtained by setting the ratio of specific heats to 
be 1.0001. We also performed simulations at the resolution of $256^3$ for 
the purposes of checking convergency and examining the 
effect of different driving schemes.  Unless explicitly stated 
otherwise, the following description of the methods and  results is based 
on the $512^3$ simulations.     

\subsection{Velocity Forcing}

To drive turbulence in our simulation, we made use of  the 
FLASH3 ``Stir" package (Benzi et al.\ 2008), which we 
heavily modified for our purposes. The flow velocity was driven by 
the forcing term, $\bs{f}(\bs{x},t)$ in eq.\ (\ref{eq:velocity}), where
the amplitude was adjusted to achieve different Mach numbers. 
In our $512^3$ simulations, we drove the turbulence with 
solenoidal modes only, i.e., $\nabla \cdot \bs{f} =0$ and took 
$\bs{f}$ to be a Gaussian random vector with an exponential 
temporal correlation.  A finite correlation 
timescale is adopted in the forcing scheme (e.g., Padoan and Nordlund 1999).  
This is  different from studies using an infinitesimal correlation 
timescale with independent forcing at each time step (e.g., Lemaster \& Stone 2009) 
or an infinite correlation timescale with a fixed driving force (e.g., Kritsuk et al.\ 2007). 
The driving scheme can be summarized as 
$\langle f_i(\bs{k}, t)f_j(\bs{k}, t') \rangle = \mathcal{P}_{\rm f}(k) 
\left( \delta_{ij} - \frac{k_i k_j}{k^2}\right) \exp \left(-\frac{t-t'}{t_{\rm f}} \right),$ 
where $t_{\rm {f} }$ is the forcing correlation timescale. The forcing wave 
numbers are chosen to be in the range $1 \le k/2\pi \le 3$ in units in which 
the domain size is 1, and each independent wave vector in 
this range is given the same forcing energy per unit time, which 
results in a quite flat forcing spectrum. This pattern was used 
in all the simulated flows, so that the shape of the forcing 
spectrum was identical for all Mach numbers. 
We also tried several $256^3$ simulations with a different driving 
scheme where we set 1/3 of the forcing energy to be in 
compressible modes.  The driving force for these flows 
was otherwise the same as for the $512^3$ simulations.

To characterize the large scale properties of the flow, we 
define a forcing length scale, $L_{\rm f}$, from the forcing spectrum, 
$L_{\rm f} = \int  \frac{2 \pi} {k} \mathcal{P}_f(k) d\bs{k}/\mathcal{P}_f(k) d\bs{k}$. 
For the forcing spectrum  chosen here, we find $L_{\rm f} =0.46$  
in units in which the domain size is 1. Lemaster \& Stone (2009) 
adopted a forcing spectrum, $\mathcal{P}_{\rm f}(k) \propto k^6 \exp(- 8 k /k_{\rm pk})$, 
that strongly peaks at a wave number $k_{\rm pk}$.  Inserting this spectrum into 
our definition for $L_{\rm f}$, we obtain that $L_{\rm f}$ is equal to $2\pi /k_{\rm pk}$, 
which is exactly the same as the characteristic flow length scale used in that 
study.  For each simulation in our study, we define a dynamical time, 
$\tau_{\rm dyn} \equiv L_{\rm f}/ v_{\rm rms}$, where 
$v_{\rm rms}$ is the 3D density-weighted rms velocity.   

We simulated turbulent flows at six different Mach numbers 
ranging from transonic to highly supersonic. At each Mach number, 
the simulation covered more than 10 dynamical 
times. After an initial phase of steady increase, the rms velocity 
saturated in a few dynamical times, and a statistically steady 
state was reached. In the steady state, the rms velocity was 
not exactly constant, but had small variations with time, with an 
(rms) amplitude of about 2\% to 5\%, depending on the 
Mach number.

Following Lemaster \& Stone (2009), we used the density-weighted (3D) 
rms velocity in our definition for the Mach number, i.e., 
$M= \langle \tilde{\rho} v^2 \rangle^{1/2}/c_{\rm s}$ where $c_{\rm s}$ 
is the sound speed. For each flow, we averaged 
the density-weighted rms velocity over about 10 dynamical times, and  
the average Mach numbers in the six flows were 0.9, 1.4, 2.1, 3.0, 4.6 and 6.1. 
If the volume-weighted rms velocity was used, the Mach number was only 
slightly larger (by about 7\% except for the $M = 0.9$ case where the two are 
essentially equal). Because all the flows were driven with the same forcing pattern, 
$L_{\rm f}$ was the same at all Mach numbers, and the dynamical timescale 
$\tau_{\rm dyn} = L_{\rm f}/v_{\rm rms}$ was proportional to $1/M$. 

\subsection{Forced and Decaying Scalars}

We studied both forced scalars and decaying scalars. 
For the forced case, we continuously added pollutants 
in the flow by the source term $S(\bs{x},t)$ in the 
advection-diffusion equation (eq.\ 1). Similar to 
the velocity driving, $S$ was taken to be a Gaussian 
variable\footnotemark\footnotetext{We point out that 
scalar sources are probably non-Gaussian in practical situations. 
However,  it is a common practice to 
set the scalar source to be Gaussian 
in studies of passive scalar physics, which is 
useful in revealing non-Gaussian scalar features 
that arise purely from mixing.}. 
with a spectrum given by $\langle S(\bs{k}, t) S(\bs{k}, t') \rangle 
= \mathcal{P}_{\rm s}(k) \exp \left(-\frac{t-t'}{t_{\rm s}} \right)$.
We considered several different forcing schemes 
for $\mathcal{P}_{\rm s}(k)$. In the first scheme, 
$\mathcal{P}_{\rm s}(k)$ was chosen to have 
the same shape as the velocity forcing spectrum,
 i.e., $\mathcal{P}_{\rm s}(k) \propto \mathcal{P}_{\rm f}(k)$. 
With exactly the same forcing pattern as the velocity, 
this scheme allows for a direct comparison of the 
dissipation timescale of the scalar variance to that of 
the kinetic energy.  As discussed below, this reveals 
important differences in the physics of scalar and 
kinetic energy dissipation, especially concerning 
the role of compressible modes and shocks. 

Like the rms velocity, the scalar variance exhibited 
temporal variations in the steady state. In order to 
reduce their amplitude, we used three independent 
scalars with the same source spectrum
with identical scalar and flow forcing. Averaging 
over the three independent scalars resulted in a 
much smaller amplitude (in the range of 2\%-5\%) 
in the temporal variations of the rms concentration than 
that of each individual scalar. 

To test the dependence of the mixing timescales on the
characteristic length scale of the scalar sources, 
we considered three other schemes in which the scalars 
were forced at larger wave numbers. In these cases, we 
chose the source wave numbers to be $3.5 \le k/2 \pi \le 4.5,$ 
$7.5  \le k/ 2\pi \le 8.5 ,$ and $16.2 \le k/ 2\pi \le 16.5,$ respectively.  Again, 
each independent mode in these ranges was given the same 
forcing power. Similar to the length scale, $L_{\rm f}$, of the 
flow driving force, we defined a source scale, $L_{\rm s} =  \int  \frac{2 \pi} {k} 
\mathcal{P}_s(k) d\bs{k}/\mathcal{P}_s(k) d\bs{k}$.   
The source length scales are $0.25$, $0.124$ and  $0.061$ 
the size of the computation box for the three cases, respectively.

In our simulations, we take the scalar source term to have 
zero mean and do not consider the evolution of the mean 
concentration, which is simply determined by the rate at which 
pollutants are injected. By definition, the concentration is positive and
bounded in the range $0 \le C \le 1$. The negative 
concentration values in our simulations should 
be understood as relative to the mean concentration. 
The amplitude of the scalar fluctuations  in 
our simulations is arbitrary (due to the arbitrarily 
chosen amplitude for the source term). This does 
not affect the study of mixing physics, because 
the advection-diffusion equation is linear with the 
concentration.    

In all the simulations, we took both $t_{\rm f}$ and $t_{\rm s}$ to be 1/4 of 
the sound crossing timescale, defined as the box size divided by the 
sound speed. We also tried different values of $t_{\rm f}$ and $t_{\rm s}$ (but with a resolution of $256^3$), 
and found that the results were not affected by the specific choice of the 
forcing/source correlation timescales.  

For the decaying case, we set an initial configuration for the 
scalar field when the turbulent flows were well developed, 
and let it evolve with no continuing sources, i.e., subsequently 
setting $S(\bs{x},t)$ to zero. For each Mach number,  
we studied four decaying scalars with different initial spectra.  
The initial scalar spectra were taken to be the same 
as the source spectra chosen for forced cases,  
i.e., the wave number ranges were set to be $ 1 \le k/2 \pi \le 3 $, $3.5 \le k/2 \pi \le 4.5,$ 
$7.5  \le k/ 2\pi \le 8.5 ,$ and $16.2 \le k/ 2\pi \le 16.5$, respectively.   

Although the main motivation for considering 
scalars forced at different scales is 
theoretical for a full understanding of mixing, 
they are also of practical interest. 
For example, if the main energy 
source for the ISM turbulence is supernova 
explosions, then, for mixing of new metals 
from SNe, the source length scale would 
be same as the flow driving scale. On the other 
hand, the scalar injection scale would be different 
from the flow forcing scale if the pollutant sources 
and the flow energy sources are different. 
An example is self-enrichment in star-forming 
clouds,  where the scalar source length scale   
may be smaller than the flow driving scale 
if the turbulent energy is mainly from the cascade 
of the interstellar turbulence at larger scales. 
The decaying case may 
be applicable to mixing in galaxies with an episodic star-formation history: 
the mixing process for elements produced by 
core-collapse SNe in between star bursts in 
these galaxies would be a decaying case.  

Finally, we point out that neither the viscous term, $\frac{1}{\rho}\partial_j \sigma_{ij}$, 
nor the diffusion term,  $\frac{1}{\rho}\partial_i (\rho \kappa \partial_i C)$, 
was explicitly included in our simulations.  Rather, kinetic energy dissipation 
and scalar dissipation were both controlled by numerical diffusion. Therefore, both 
the viscous dissipation scale and the diffusion scale are expected to be 
around the resolution scale. The effective viscosity and diffusivity are thus similar, 
with the effective Schimdt number, $Sc \approx 1$, and the 
Peclet number about equal to the Reynolds number. For the 
case with identical scalar and flow forcing, the convective scale range  
coincides with the inertial range, while the scalars forced at larger wave 
numbers has a smaller convective range than the inertial range.  

\subsection {Calculation of the Timescales}

To calculate the scalar dissipation timescale, $\tau_{\rm c}$, we 
need the scalar dissipation rate, $\epsilon_{\rm c}$, which
could not be directly computed  in our simulations because the 
dissipation was controlled by numerical diffusion. However, 
in the case of forced scalars, after the scalar fluctuations reach a statistically stationary state, 
we have a balance between the dissipation term and the source term in eq.\ (\ref{eq:Cvariance}),  
i.e., $\epsilon_{\rm c} = \epsilon_{\rm s}\equiv \langle \tilde{\rho} S C \rangle$. 
The computation of the source term is 
straightforward, and we calculated the scalar dissipation timescale as 
$\tau_{\rm c} =\langle \tilde{\rho} C^2 \rangle/\langle \tilde{\rho} S C \rangle$. 
For the scheme with identical scalar and flow forcing,  we used 
the average of the source term over the three independent scalars.  
 
Similarly, the viscous dissipation of kinetic energy was obtained using 
energy balance, $\bar{\epsilon}_{\rm f} -\bar{\epsilon}_{\rm p} - \bar{\epsilon}_{\nu} 
=0$, in the statistically stationary state. The energy injection by the driving force 
per unit time, $\bar{\epsilon}_{\rm f}$, and the contribution from pdV work, $\bar{\epsilon}_{\rm p}$ were 
calculated by $\langle \tilde{\rho} f_i v_i \rangle$ and $\bar \rho^{-1} \langle p \partial_i v_i \rangle$ in eq.\
 (\ref{eq:keperunitmass}), respectively. In the stationary state, this allowed us to evaluate the 
 viscous dissipation timescale as $\tau_{\rm diss} = \langle \frac{1}{2}\tilde{\rho} v^2 \rangle/ \left( \bar{\epsilon}_{\rm f} -\bar{\epsilon}_{\rm p} \right)$,
and  the timescale for the overall kinetic energy loss including the contribution from the pdV work
as $\tau_{\rm v} = \langle \frac{1}{2}\tilde{\rho} v^2 \rangle/\bar{\epsilon}_{\rm f}$.  

The kinetic energy loss by both the pdV work and the viscous dissipation was 
stored as thermal energy. We point out that the isothermal 
equation of state adopted in our simulations was used to imitate a roughly 
constant temperature due to the effect of efficient radiative cooling in 
interstellar clouds. In nature, the heat converted from kinetic 
energy would be radiated away. 
 
We find that, if the forcing correlation timescale, $t_{\rm{f}}$, was much smaller 
than the typical timescale for the kinetic energy loss to thermal energy, $\tau_{\rm v}$, 
the energy injection rate goes like $\bar{\epsilon}_{\rm f} \simeq  t_{\rm f}  \int  \mathcal{P}_{\rm f}(k) d{\bs k}$. 
On the other hand, if $t_{\rm{f}}$ is larger than $\tau_{\rm v},$ as is the case for our choice of $\tau_{\rm{f}}$,  
$\bar{\epsilon}_{\rm f} \propto \tau_{\rm{v}} \int  \mathcal{P}_f(k) d{\bs k}$ because only the forcing 
energy within a timescale $\tau_{\rm v}$ remains as kinetic energy in the flow.  
The same applies to the variance input from scalar sources, i.e.,  $\bar{\epsilon}_{\rm s} 
\simeq  t_{\rm s}  \int  \mathcal{P}_{\rm s} (k) d{\bs k}$ if $t_{\rm s} \ll \tau_{\rm c}$ or 
$\simeq \tau_{\rm c}  \int  \mathcal{P}_{\rm s} (k) d{\bs k}$ for $t_{\rm s} \gg \tau_{\rm c}$. 

Similar to the velocity and scalar variances, the other quantities needed 
for the calculation of timescales also show temporal variations to different 
degrees. These quantities include the kinetic energy injection by the 
driving force, the pdV work, and the scalar variance input by the sources. 
In our calculations, we used the temporal average of all these quantities 
over about 10 dynamical times. Comparing with results from 256$^3$ 
simulations for the corresponding flows, we find that the timescales have 
converged at our resolution of 512$^3$ zones (see also Lemaster \& Stone 2009). 

\begin{figure*}
%\epsscale{1}
\centerline{\includegraphics[height=16cm]{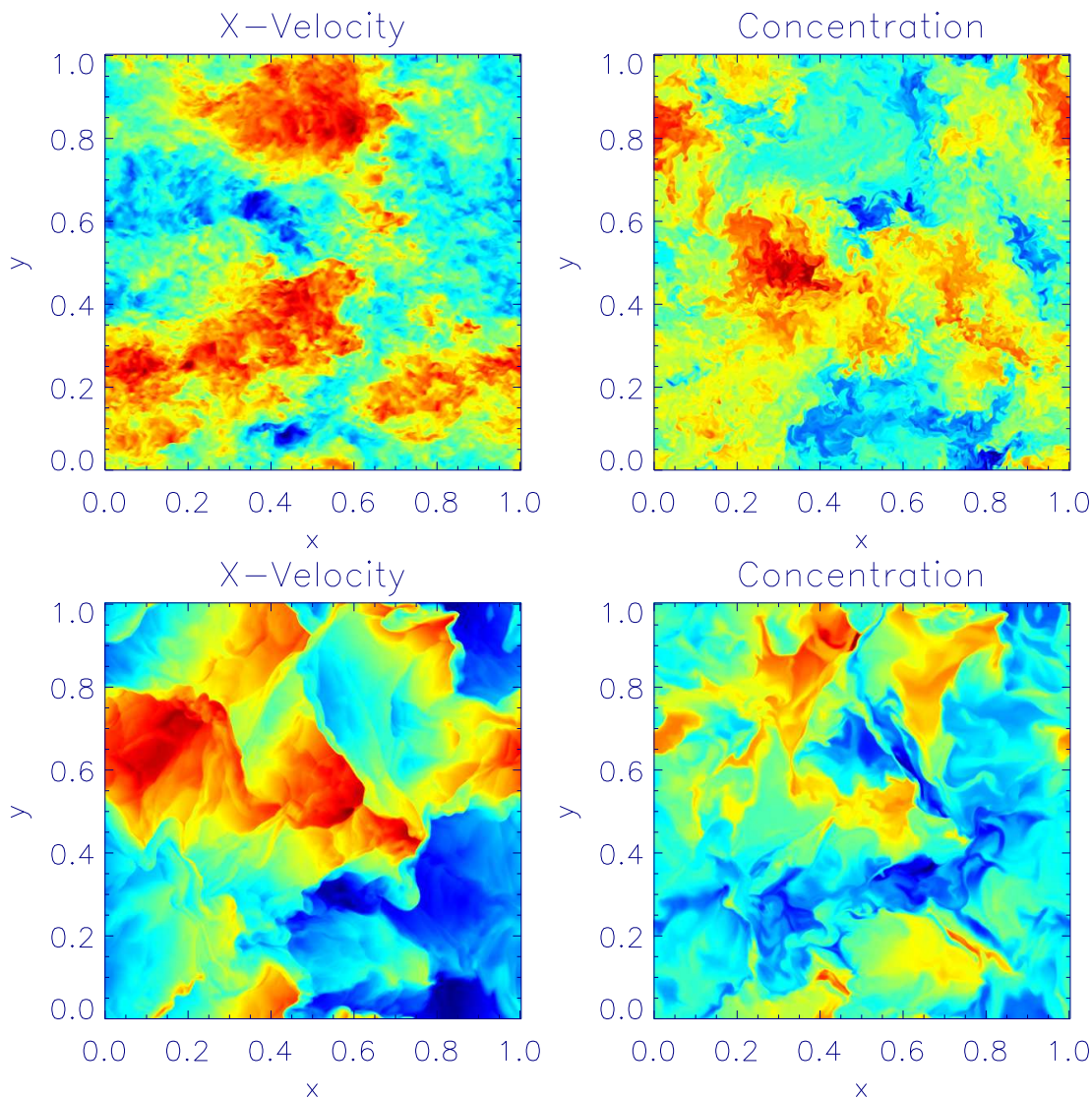}}
\vspace{-0.8cm}
\caption{The flow velocity in X direction (left) and the 
scalar field (right) on a slice of the simulation grid, with a linear  
color scale going from the minimum value  (blue) to the 
maximum value (red) on the slice. 
The top two panels are for the $M=0.9$ flow at  
$t= 8.0 \tau_{\rm dyn} $, and the bottom two panels 
correspond to the $M=6.1$ flow at $5.6 \tau_{\rm dyn}$.}
\vspace{0.2cm}
\label{fig:slice}
\end{figure*}

\section{Results}    

\subsection{The Turbulent Flows}
In the left panels of Fig.\ \ref {fig:slice}, we show the X-component of 
the velocity field on a slice (X-Y plane) 
of the simulation grid  at one snapshot for two Mach numbers, $M=0.9$ and $M=6.1$.
The snapshots are taken at $t= 8 \tau_{\rm dyn}$ and 
$5.6 \tau_{\rm dyn}$ for $M=0.9$ and $M=6.1$, respectively.
The concentration field shown here is from a scalar forced 
at the flow driving scale. The velocity field exhibits extremely 
different features in these two limits. In the  $M=0.9$ case, the flow contains a 
large number of vortices, which persist down to small
scales. In the $M=6.1$ case, on the other hand, the field appears 
to be dominated by large-scale expansions and compressions, 
and the most prominent structures are shock fronts. 
However, we point out that, although not clearly visible in Fig.\ \ref {fig:slice}, 
a significant fraction of the kinetic energy is contained in 
solenoidal modes even at $M=6.1$. For example, strong 
shears and vortices are usually found behind narrow 
postshock regions. As we will show later, the solenoidal 
modes play important roles in both energy 
dissipation and mixing at all Mach numbers.

In Table 1, we list the overall properties of the flow as a 
function of Mach number. In the 2nd column, we give the 
ratio of the density-weighted divergence variance and vorticity variance, 
$\langle \tilde{\rho} (\nabla \cdot \bs{v})^2 \rangle/\langle \tilde{\rho} (\nabla \times \bs{v})^2 \rangle$. 
This ratio characterizes the contribution of compressible modes 
to  viscous dissipation relative to solenoidal modes,
and it is smaller than 1 even for $M=6.1$, suggesting that the 
majority of kinetic energy is dissipated by solenoidal 
modes even in highly supersonic flows (Kritsuk et al.\ 2007).  
However, we point out that this ratio is only a rough 
estimate for the relative contribution from the two modes, 
as it is not clear how the kinetic energy is dissipated by 
numerical diffusion by the high-order PPM solver.
In the 3rd column, we also give the same ratios, but 
with no density weighting. For both cases, the 
ratio increases with the Mach number at small $M$. 
The density-weighted ratio saturates at $M \approx 3$, 
while the volume weighted one saturates at a little larger $M,$ 
and the density-weighted ratios are all smaller because of an 
anti-correlation between the density and divergence.

The 4th column lists the fraction of kinetic energy in 
compressible modes. We computed this fraction by 
decomposing the energy spectrum into a solenoidal 
part and a potential part and comparing the 
kinetic energy contained in each. Although the 
flow is driven by solenoidal modes only, a finite fraction 
of kinetic energy appears in the compressible modes, which 
are converted from the solenoidal modes.  The 
fraction increases from 5\% at $M=0.9$ to 20\% at $M=6.1$ 
in our simulations.  We note that these values are not universal, 
but depend on the ratio of the compressible modes to 
the solenoidal modes in the driving term, as discussed below.

We are particularly interested in the velocity fluctuations 
in the inertial range, which regulate the scalar cascade 
to the diffusion scale.  For scalars forced at the flow driving scale,  
the cascade starts at the same scale as the 
energy cascade, i.e., at $k/2 \pi \approx 4$.   
Thus we calculated the kinetic energy contained in 
the compressible modes and the solenoidal modes
in the inertial range by integrating the potential and 
solenoidal spectra over all wave numbers $k/2 \pi \ge 4,$ 
which also includes the negligible contribution from the 
dissipation range. The 5th column in Table 1 
gives the fraction of kinetic energy in compressible 
modes at $k/2 \pi \ge 4$, which increases at small 
$M$ and saturates at about 1/3 for $M \gsim 3$. 
The fraction of 1/3 implies an energy equipartition 
between the solenoidal modes and the potential 
modes in the inertial range, which is reached at 
$M \approx 3$. We find the slope of the compressible 
spectrum is close to that of the solenoidal 
spectrum in the inertial range, meaning that a 
equipartition is established at each wave number 
in the inertial range for $M \approx 3$. We find this
equipartition is universal regardless the 
compressible fraction in the flow driving (see below).  

\begin{table*}
\begin{center}
\caption{Flow properties at different Mach numbers}
\label{tbl-1}
\begin{tabular}{cccccc}
\tableline\tableline
$M$ & $\langle \tilde{\rho} (\nabla \cdot \bs{v})^2 \rangle/ \langle \tilde{\rho} (\nabla \times \bs{v})^2 \rangle$ 
& $ \langle (\nabla \cdot \bs{v})^2 \rangle/\langle (\nabla \times \bs{v})^2 \rangle$ & $f_{\rm comp}$ & $f_{\rm comp} (k/2\pi \ge 4)$ & $f_{\rm pdV}$ \\
\tableline
0.9 &  0.06 & 0.06& 0.05 &0.07 & 0.15  \\
1.4 &  0.19 & 0.20& 0.10 &0.16 & 0.30  \\
2.1 &  0.41 & 0.49& 0.15 &0.25 & 0.35  \\
3.0 &  0.52 & 0.75& 0.17 &0.30 & 0.30  \\
4.6 &  0.55 & 0.88& 0.18 &0.33 & 0.23 \\
6.1 &  0.54 & 0.86& 0.19 &0.32 & 0.17 \\
\tableline
\end{tabular}
\end{center}
\end{table*}
  
Note that the fractions given in columns 4 and  5 do not
account for density-weighting, which cannot be perfectly 
applied to the fraction of kinetic energy in compressible 
modes at large $M$.  Although 
the velocity field can be decomposed into a solenoidal part, 
$\bs{v}_{\rm s}$, and a potential part, $\bs{v}_{\rm p}$, 
the density-weighed ratio 
$\langle \tilde{\rho} v_{\rm p}^2 \rangle/(\langle \tilde{\rho} v_{\rm p}^2 \rangle + \langle \tilde{\rho} v_{\rm s}^2 \rangle)$
cannot be interpreted as the energy fraction at high Mach numbers.
This is because the denominator 
is not equal to the total energy, which is given by 
$\langle \tilde{\rho} v_{\rm p}^2 \rangle + 2\langle \tilde{\rho} \bs{v}_{\rm p} \cdot \bs{v}_{\rm s}  \rangle + \langle \tilde{\rho} v_{\rm s}^2 \rangle$.  
In fact, we find that the density-weighted correlation term $\langle \tilde{\rho} \bs{v}_{\rm p} \cdot \bs{v}_{\rm s}  \rangle$ is 
not zero. Rather, it is negative, decreases with increasing $M,$ 
and becomes significant at $M\gsim 2-3$. Therefore, 
density-weighting for the fraction of energy in the compressible modes cannot 
be well defined at large Mach numbers. However, there is no such 
problem for the volumed-weighted fraction because it can be 
shown that $\langle \bs{v}_{\rm p} \cdot \bs{v}_{\rm s}  \rangle$ is 
exactly zero in the homogeneous case, and we thus  only 
show the volume-weighted fraction in Table 1.

The last column in Table 1 lists the ratio of the kinetic 
energy loss by pdV work to the total kinetic energy loss. This will 
be discussed in detail in \S 4.3.1.

To test whether and how the flow properties listed in Table 1 
depend on the flow driving, we performed a set of ($256^3$) 
simulations covering a similar range of Mach numbers 
but with 1/3 forcing energy contained in the compressible modes.   
We found that, with this different driving scheme, most 
quantities listed in Table 1 remain essentially unchanged,  
except  the overall fraction of energy in the 
compressible modes (column 4).  This is expected 
since the overall compressible fraction 
depends on the driving scales, which contain most of 
the kinetic energy, while the rest of the 
quantities depend on the statistics at 
smaller scales, which should 
be driving-independent.        

The change in the overall compressible 
fraction is slight for $M \lsim 2$,  
only a few percent larger than the case with solenoidal driving,  
but there is a significant increase in the fraction 
at $M \gsim 3$. With the compressible driving 
scheme, the fraction saturates at $\approx$24\% for $M \gsim 3$,  
considerably larger than $\approx$ 18\% given in 
Table 1. We note that the overall compressible 
fraction in the flow turns out to be smaller than 
that (1/3) in the driving force at all Mach numbers, indicating 
a preferential conversion from compressible
modes to solenoidal modes at the driving scales. 
This is different from the case with solenoidal 
driving where a fraction of kinetic energy in  
solenoidal modes is converted to compressible 
modes.  
   
Of particular interest is that, in the inertial range, 
the equipartition between solenoidal and compressible modes 
is also found in the $M \gsim 3$ 
flows that are forced with 1/3 driving energy 
in compressible modes, suggesting that this  
equipartition at high Mach numbers is universal.  
On the other hand, no equipartition between  
compressible and solenoidal modes exists at the 
driving scales even in the case where the 
compressible fraction in the driving force is 
set to the equipartition value of 1/3. 
 
\subsection{The Concentration Field}

In the right panels of Fig.\ \ref {fig:slice}, we show 
the concentration on slices taken from simulations 
with Mach numbers $M=0.9$ and $M=6.1$. As was 
the case for the velocity field, there are large  
differences between the two runs. At $M=0.9,$ the scalar 
field shows numerous small-scale structures, which 
are clearly produced by stretching and shear in 
vortices. There are  many structures with sharp 
concentration contrasts, which are usually referred to 
as ``cliffs and ramps" in the literature for passive 
scalar in incompressible turbulence 
(e.g., Shraiman \& Siggia 2000).  These cliffs and 
ramps are produced by turbulent stretching,  
which rearranges the scalar gradients that initially 
exist only at large scales and brings fluid 
elements with very different concentration levels 
next to each other. The ``cliffs and ramps" have large 
scalar gradients, and are thus strongly dissipative.    

On the other hand, the scalar field at $M=6.1$ 
primarily reflects expansions and compressions 
in the flow, similar to the velocity field at this Mach 
number. Also like the velocity case, the visual 
impression of the figure is biased toward compressible 
modes, and does not sufficiently reflect the effect 
of solenoidal modes on the scalar field, which we will show is 
dominant for mixing.

An interesting property of this 
scalar field is the existence of scalar 
``edges" that coincide 
with the locations of velocity shocks. Unlike the 
cliff and ramp structures in the $M=0.9$ case,
these edges are produced not by stretching, 
but rather by compression by shocks, which squeeze 
the concentration profile and bring spatially well 
separated regions with different concentrations 
close to each other. In this way, the length scale 
of the concentration profile is reduced by a 
factor about equal to the density jump across 
the shock, and the concentration gradient 
is amplified by the same factor. 

Consider for example a shock with a density jump 
factor of 20 in our simulation. Squeezing by this 
shock could bring two fluid elements in 
the pre-shock region with a distance of 
20 times resolution scales to a separation 
of one computation cell. If the two elements 
have quite different concentrations, then, 
when they pass the shock, one would see a 
significant concentration change across the 
shock front, i.e., a scalar edge.   
However, if the fluid elements in the preshock 
region have the same concentration, i.e., if the 
preshock scalar profile is uniform, no scalar edge 
would form. This explains why scalar edges 
are not found at every velocity shock front.  

Gradient amplification is not limited to shock 
fronts, but instead occurs in any region that has 
been compressed. For example, in post shock regions, 
the scalar field is also squeezed by compression, 
and thus large gradients are expected 
behind shocks. In fact, in Fig.\ 1, we see there 
is usually a second sharp concentration change 
right behind a scalar edge, which may correspond 
to the amplified gradient in the postshock region. 
However, extended regions with strong gradient 
amplification are not observed behind the edges, 
perhaps because postshock regions 
are narrow. Also these regions are subject to 
rapid mixing because the scalar gradients 
there are amplified by both compression 
and strong shears and vortices, which usually 
exist behind shocks.  

As compression events give rise to 
larger scalar gradients, they increase the 
mixing efficiency, which depends on the
generation of large-gradient structures. 
On the other hand, the 
compressible modes also include expansions, 
which reduce the scalar gradients, and thus 
decrease the mixing rate. 
Thus, it is not clear if the presence
of compressible modes leads to a
net  increase of the mixing efficiency.  
This will be 
addressed in more detail in \S 5, where we also 
compare the efficiency of the compressible modes 
relative to the solenoidal modes in amplifying the scalar gradients. 

Note that there is a fundamental difference between 
scalar edges and velocity shocks. Although 
the scalar edges in Fig.\ 1 appear to be 
discontinuous, they are actually 
continuous. This can be shown by considering 
the concentration change across a shock 
front in the limit of infinitesimal shock thickness. 
From the formation mechanism 
of the scalar edges, the concentration change 
over a shock front can be roughly estimated by 
the product of three factors: 
the preshock gradient, the density jump factor 
(i.e., the squeezing factor) and the shock thickness. 
Therefore, when the shock thickness approaches 
zero (corresponding to decreasing viscosity), 
the concentration change across the shock front 
would also approach zero, since the density jump 
factor would remain constant and finite as the 
shock width decreases to infinitesimal. This 
means that the concentration is essentially 
continuous, unlike the velocity shocks, which are exact 
discontinuities in the limit of infinitesimal 
viscosity. This argument is confirmed 
by simulations for 1D supersonic flows 
at very high resolutions.  

The continuity of the scalar field across a 
shock of infinitesimal width is also expected 
from the conservation of the tracer mass.  
Mass conservation implies that the 
tracer density has the same jump as the 
flow density across a shock, and thus the 
concentration, which is the ratio of the 
two densities, would be continuous across 
the front.  

The reason the edges appear to be 
discontinuous in our simulation is due to 
the limited resolution. A finite and significant 
shock width in our simulations means that 
the materials across the shock front 
could come from well separated regions 
(especially for strong shocks), making 
a considerable scalar change across the 
shock front possible. With increasing 
resolution, we would expect the scalar structures 
around the shocks appear to be more continuous, while 
the shocks would clearly remain discontinuous.  
If a shock had an infinitesimal width, its effect 
on the scalar field is just to amplify the scalar 
gradient in the postshock region.

The difference between the scalar edges and 
the velocity shocks has an interesting implication. 
Every shock is intrinsically a strong dissipative 
structure for kinetic energy, but it is not true that 
every shock produces a strong dissipative 
scalar structure, considering that a scalar edge 
is essentially not a discontinuity. In fact, most 
shocks cannot produce scalar structures that 
dissipate scalar fluctuations at the same level as shocks for 
energy dissipation. To produce such a structure, 
a shock needs to reduce the length scale 
of a scalar structure close to the diffusion scale. 
Whether a shock can give rise to this scale reduction 
depends on the shock strength, which determines the 
squeezing factor across the shock. We will show that strong 
shocks capable of such a scale reduction are very 
rare in \S 5, and this point is responsible for the 
difference in the roles of shocks in kinetic energy 
dissipation and in scalar dissipation. 

\subsection{Timescales}

\subsubsection{Kinetic Energy Dissipation}

\begin{figure}
\centerline{\includegraphics[height=7.5cm]{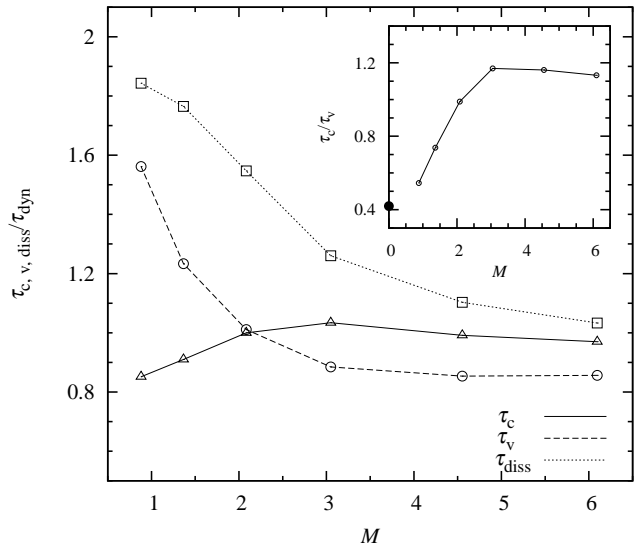}}
\caption{The scalar dissipation timescale, $\tau_{\rm c}$, the 
viscous dissipation timescale, $\tau_{\rm diss}$, and the timescale
for the overall kinetic energy loss, $\tau_{\rm v}$,  as functions of the 
Mach number. The timescales are normalized to the dynamical time 
$\tau_{\rm dyn}$, such that the behavior of the timescales with 
$M$ depends mainly on the effect of 
compressible modes relative to the solenoidal modes. The inset 
shows the ratio of $\tau_{\rm c}$ to $\tau_{\rm v}$, and 
the filled circle is from the simulation result by Donzis et al.\ (2005) for 
incompressible turbulence.}
\label{fig:timescales}
\end{figure}

In Fig.\ \ref{fig:timescales}, we plot the timescales defined 
in \S 3.1 and \S 3.2 as a function of Mach number,
normalized to the dynamical time, $\tau_{\rm dyn}$. 
The behavior of the normalized timescales as a function of 
$M$ is expected to essentially reflect the contribution from 
compressible modes relative to solenoidal 
modes.    

The dotted line in Fig.\ 2 shows the normalized 
timescale for the kinetic energy loss by 
viscous dissipation, $\tau_{\rm diss}/\tau_{\rm dyn}$. 
This timescale steadily decreases as $M$ increases from 0.9 to 3, 
and shocks become stronger and more frequent.  Shocks 
are strong dissipative structures for kinetic energy, and thus 
compressible modes provide an extra fast channel for the viscous 
dissipation of kinetic energy, which does not exist in incompressible 
flows. The steady decrease in the $\tau_{\rm diss}/\tau_{\rm dyn}$ 
curve with $M$ for $M \lsim 3$ parallels the increase in the 
ratio of the divergence variance to the vorticity variance 
listed in Table 1 with and without density weighting.  
The ratio becomes roughly constant for $M \gsim 3$, 
corresponding to the slower decrease of $\tau_{\rm diss}/\tau_{\rm dyn}$ 
in that range of $M$. This suggests that the faster viscous dissipation 
at larger $M$ may be primarily due to a larger  
contribution from the compressible modes.  
Overall the density-weighted divergence to 
vorticity variance ratio increases by about 50\% as 
the Mach number increases from $M=0.9$ to 6.1,  
generally consistent with the decrease in the 
viscous dissipation timescale by a factor of 1.8.

The dashed curve in Fig.\ 2 is $\tau_{\rm v}/\tau_{\rm dyn}$, 
the normalized timescale for the overall kinetic energy loss 
by both pdV work and viscous dissipation. This 
timescale is significantly smaller than $\tau_{\rm diss}$ 
because the pdV work gives a considerable contribution 
to the kinetic energy loss in the chosen range of $M$. 

The fraction of kinetic energy that is converted to 
heat by pdV work is given in the 6th column of Table 1, 
which shows that the fraction peaks at the intermediate 
Mach number, $M \simeq 2$. At low Mach numbers, 
increasing $M$ allows more compressible 
modes to appear and larger $| \nabla \cdot \bs{v} |$ to be 
available, leading to a rapid increase in the rate of 
pdV work. On the other hand, the importance of 
pdV work in converting kinetic energy into heat 
{\em decreases} with $M$ for $M \gsim 2$. 
At these large Mach numbers, the amplitude of 
the divergence scales linearly with $M$, corresponding 
to the saturation of the fraction of energy in 
compressible modes (see Table 1) and  
the increase in the amplitude of pressure fluctuations is 
slower than $\propto M$. Therefore the conversion 
rate by pdV work does not increase faster than $\propto M^2$. 
 Since the viscous dissipation rate increases as $\propto 
 v_{\rm rms}^3$, this means that 
 the fraction of energy loss by pdV work decreases 
 at large $M$.  

This trend is also expected from another 
perspective. The viscous dissipation rate by 
compressible modes is a second-order function 
of the velocity divergence and increases with 
$M$ faster than the pdV work, which scales 
linearly with the divergence. Thus, as $M$ 
increases above 2, the fraction of kinetic energy 
loss by pdV work starts to decrease as the 
viscous dissipation rate by compressible modes 
becomes a significant fraction of the total viscous dissipation rate. 

The increase in the fraction of kinetic 
energy loss by pdV work for $M \lsim 2$ makes the 
$\tau_{\rm v}/\tau_{\rm dyn}$ curve a little steeper than 
the curve for,  $\tau_{\rm diss}/\tau_{\rm dyn}$, the normalized 
timescale for viscous dissipation only.  On the other hand, 
the $\tau_{\rm v}/\tau_{\rm dyn}$ curve is 
flatter than the $\tau_{\rm diss}/\tau_{\rm dyn}$ curve for 
$M \gsim 3$ due to the 
decrease in the pdV work fraction, which  
happens to cancel the weak increase in the viscous dissipation 
rate.  The overall kinetic energy loss rate in highly supersonic 
flows ($M>3$) is faster than in the transonic case with 
$M =0.9$ also by a factor of $\approx 1.8$.

Our simulations with 1/3 forcing energy in 
compressible modes give essentially the same 
results for the energy dissipation timescales 
as those found from pure solenoidal driving. 
Clearly, the energy dissipation occurs at small 
scales and basically depends on the cascade in 
the inertial range. As discussed earlier, the 
statistical measures at small scales listed in Table 1 
(i.e., except the 4th column) are the same in the 
two different driving schemes, and thus the 
timescales for kinetic energy dissipation 
are expected to be also driving-independent.  

We note that, although our forcing scheme is very different 
from that in Lemaster \& Stone (2009), the timescale 
$\tau_{\rm v}$ from our simulations is in good agreement 
with the results given in their Table 1 
(the column for $t_{\rm diss}/t_{\rm f}$).

\subsubsection{Scalar Dissipation}

The solid curve in Fig.\ \ref{fig:timescales} shows the 
timescale for the scalar dissipation, $t_{\rm c},$ focusing 
on the case in which the source pattern is identical to the 
flow driving. The behavior of the scalar dissipation timescale 
is completely different from that of the  kinetic energy, although 
all the scalar fields are driven in exactly the same manner 
as the velocity fields. Unlike the kinetic energy dissipation 
timescale, the normalized timescale for scalar dissipation, 
$\tau_{\rm c}/\tau_{\rm dyn}$, increases by about 20\% as $M$ 
goes from $0.9$ to $3$. This increase in the mixing 
timescale parallels the decrease  
in the fraction of kinetic energy in solenoidal modes 
in the inertial range, which goes from 93\% to 70\% 
in the same range of $M$ (see Table 1).  

This indicates that the existence of more compressible modes reduces 
the mixing efficiency.  There are two effects that may contribute 
to the increase of the mixing timescale. First,  as shown in 
more detail below (\S 5), the compressible modes 
are less efficient than the solenoidal modes in producing 
structures small enough for the molecular diffusivity to homogenize. 
Therefore, with more kinetic energy contained in compressible modes, 
the overall efficiency for the production of small scale structures 
decreases, resulting in an increase in the scalar dissipation timescale. The second effect 
is from the steepening of the velocity structure function 
with $M$ (see \S 4.4). As discussed  in \S 4.3.3, a steeper 
velocity structure function suggests the time spent on cascade steps 
at small scales is relatively larger, and this may result in a slight increase 
of the overall timescale for the entire cascade. The second effect may be viewed as an indirect consequence of 
more energy in compressible modes at larger $M$, 
which tends to make the structure function steeper  (\S 4.4).

\begin{table*}
\begin{center}
\caption{Energy dissipation and mixing timescales}
\label{tbl-2}
\begin{tabular}{ccccccc}
\tableline\tableline
$M$ & $\tau_{\rm diss}/\tau_{\rm dyn}$ &  $\tau_{\rm v}/\tau_{\rm dyn}$ & \multicolumn{4}{c}{$\tau_{\rm c}/\tau_{\rm dyn}$} \\\cline{4-7}
           &                                                      &                                                     
           & {\small $1 \le k/2\pi \le 3$ } &  {\small $3.5 \le k/2\pi \le 4.5$}  &  {\small $7.5 \le k/2\pi  \le 8.5$}  &  {\small $16.2 \le k/2\pi \le 16.5$}\\
\tableline
0.9 &  1.84 &  1.56 &0.85 & 0.57 &  0.37  &  0.23  \\
1.4 &  1.76 &  1.23 &0.91 & 0.63 &  0.41  &  0.27  \\
2.1 &  1.55 &  1.01 &1.00 & 0.73 &  0.50  &  0.33  \\
3.0 &  1.26 &  0.88 &1.03 & 0.77 &  0.55  &  0.37  \\
4.6 &  1.10 &  0.85 &0.99 & 0.75 &  0.55  &  0.37  \\
6.1 &  1.03 &  0.86 &0.97 & 0.74&   0.54  &  0.38  \\
\tableline
\end{tabular}
\end{center}
\end{table*}

Above $M=3$, $\tau_{\rm c}/\tau_{\rm dyn}$ is 
almost constant. Again, this is consistent with the fraction 
of kinetic energy in solenoidal modes, which is 
constant at 66\% in the inertial range due to energy 
equipartition. We note that there a slight decrease in 
the mixing timescale curve in Fig.\ \ref{fig:timescales} as $M$ 
goes from 3 to 6.1. This is related to the fact that, although 
the energy fraction in compressible modes in the inertial range is constant  
in this range of Mach numbers, the probability of strong 
compression events increases with $M$, as can be seen 
from the increase in the amplitude of density fluctuations. 
As discussed in \S 5, the 
compressible modes do contribute to the amplification 
of the scalar gradient, although the contribution is tiny 
in comparison to the solenoidal modes. At the current 
resolution (512$^3$), the contribution to the gradient 
amplification by compressible modes is a few percent 
at $M = 6.1$ (see \S 5), while it is completely 
negligible at $M=3$. This is responsible for the slight 
decrease in the mixing timescale in this range of Mach numbers  
\footnotemark\footnotetext{ 
In \S4.4, we find that there is a slight 
steepening in the velocity structure function 
as $M$ goes from $3$ to $6.1$, which may 
tend to give a slight increase in the mixing 
timescale.  The actual decrease in the 
normalized mixing timescales in this 
range of $M$ means that this effect is weaker 
than the opposite effect due to the increasing 
contribution from compressible 
modes to gradient amplification discussed here.}.  

Like the timescale for kinetic energy dissipation, 
the mixing timescales in our 256$^3$ simulations 
with 1/3 forcing energy in the compressible modes 
are the same as those given in Fig.\ 2 from 
pure solenoidal driving.  Although one may suspect 
the larger overall compressible fraction from the 
compressible driving scheme could affect the 
mixing efficiency, that is not true because the 
scalar cascade is controlled by the velocity 
fluctuations in the inertial range.  Because 
the compressible energy fraction in the inertial 
range is independent of the compressible fraction in the 
driving force, the mixing timescale remains the same.   

The inset in Fig.\ \ref{fig:timescales} shows the ratio of the scalar 
and kinetic dissipation times, $\tau_{\rm c}/ \tau_{\rm v}$. Because 
of the contrary behavior of these two timescales, their ratio 
increases by about a factor of 2.6 from incompressible turbulence 
to $M=3$, and saturates at about 1.1 for large Mach numbers. 
The filled circle is taken from the simulation results by Donzis 
et al.\ (2005) for incompressible turbulence and is 
consistent with our results.

In summary, we find that the role of compressible 
modes is completely different for mixing than for
kinetic energy dissipation.  By providing an additional 
channel for energy dissipation,  compressible modes 
lead to faster energy loss at larger Mach numbers. 
On the other hand, mixing tends to be relatively slower 
if more kinetic energy is contained in compressible modes, 
suggesting that they are not efficient in producing small 
scalar structures. However, since the majority of kinetic 
energy is contained in solenoidal modes even in flows at 
very large Mach numbers, the mixing rate changes 
only weakly at increasing 
Mach numbers. Thus, the mixing timescale only increases 
by 20\% from $M=0.9$ to $M=3$ and always remains 
comparable to the dynamical timescale.

In fact, although their detailed behaviors as a function of 
$M$ are different, the overall timescales for the 
scalar dissipation and for the kinetic energy loss are 
similar at all Mach numbers. This suggests that the 
production of scalar structures at small scales occurs 
through a cascade process similar to that of kinetic energy.  
In this case, because the scalar and velocities cascades 
start at the same physical scale, their decay timescales are 
both expected to be close to the turnover time of 
the largest turbulent eddies, which is essentially the 
dynamical timescale. This is exactly what we find in 
our simulations. 

\subsubsection{Scale Dependence of  the Mixing Timescale}

\begin{figure}
\centerline{\includegraphics[height=7cm]{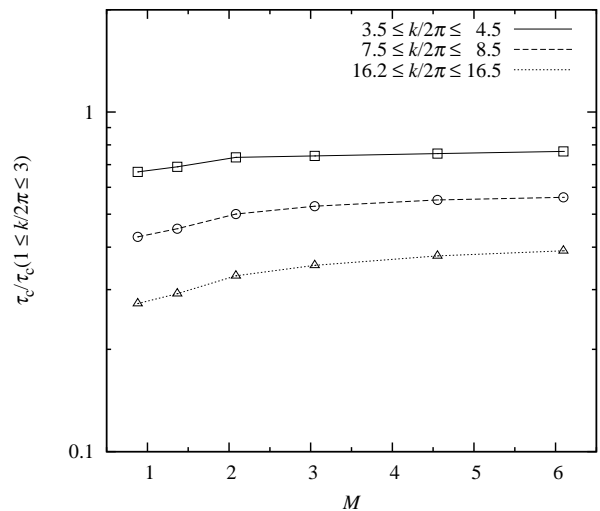}}
\caption{The dependence of the mixing timescale, $\tau_{\rm c}$, 
on the scalar source spectrum. The timescales are normalized 
to $\tau_{\rm c}$ in the case in which the scalar and the flow are driven 
at the same scales. The three lines are results from different source 
spectra, whose characteristic length scales, $L_{\rm s}$, are, 
respectively, 0.25 (solid), 0.124 (dashed), and 0.061 (dotted) 
times the computation box size. }
\vspace{0.1in}
\label{fig:timewithscale}
\end{figure}

In Fig.\ \ref{fig:timewithscale},
we show the dependence of $\tau_{\rm c}$ 
on the characteristic length scale of scalar sources. 
The three curves are for three scalars forced in 
different wave number ranges:    
$3.5  \le k/2 \pi \le 4.5,$ $7.5 \le k/ 2 \pi \le 8.5,$ and $16.2 \le k/ 2\pi \le 16.5$. 
The three mixing timescales are normalized 
to that of  the scalar forced  in the same 
wave number range as the flow driving. The corresponding mixing 
timescales normalized to the dynamical time are given in 
Table 2.

The results in Fig.\ \ref{fig:timewithscale} suggest that the 
mixing timescale is proportional to the turnover time of 
eddies at the length scale of the scalar sources. 
At all Mach numbers, the mixing timescales decrease with 
decreasing source length scale. This is because the turnover 
timescale in smaller eddies is shorter, and thus, starting 
from a smaller source scale, the cascade to the diffusion 
scale is faster. Fig.\ \ref{fig:timewithscale} also shows that, 
with increasing Mach numbers, the scale dependence 
becomes weaker,  which corresponds to the steepening 
of the velocity structure functions (see \S 4.4). 
At larger Mach number, the velocity field has relatively less 
power at small scales, and thus the increase in the scalar 
cascade toward small scales becomes slower, because the 
decrease in the eddy turnover time is slower. 

Quantitatively, we find that the scaling of mixing 
timescales with the source size is consistent with that of 
the eddy turnover time with the eddy size.  We consider the 
two forced scalars with  source length scales 
$L_{\rm s} =0.124$ and 0.061, which lie in the intertial range.
Roughly fitting the mixing timescales for these two 
scalars with a power law, $\tau_{\rm c} \propto  L_{\rm s}^{\alpha}$, 
gives $\alpha= 0.64, 0.62, 0.60, 0.57, 0.54, 0.52$ for Mach 
numbers from 0.9 to 6.1. As we will see below, these numbers 
are quite close to the scaling exponents of the turnover time,  
$l/\delta v(l)$, as a function of the eddy size $l$, using the 
inertial-range scaling of $\delta v(l).$ 
We thus conclude that the mixing timescale is 
essentially given by the turnover time of eddies at 
the scalar injection scale.

\subsection{Structure Functions and Power Spectra}

\begin{figure*}
%\epsscale{1.}
%\plotone{f4.eps} 
\centerline{\includegraphics[height=6cm]{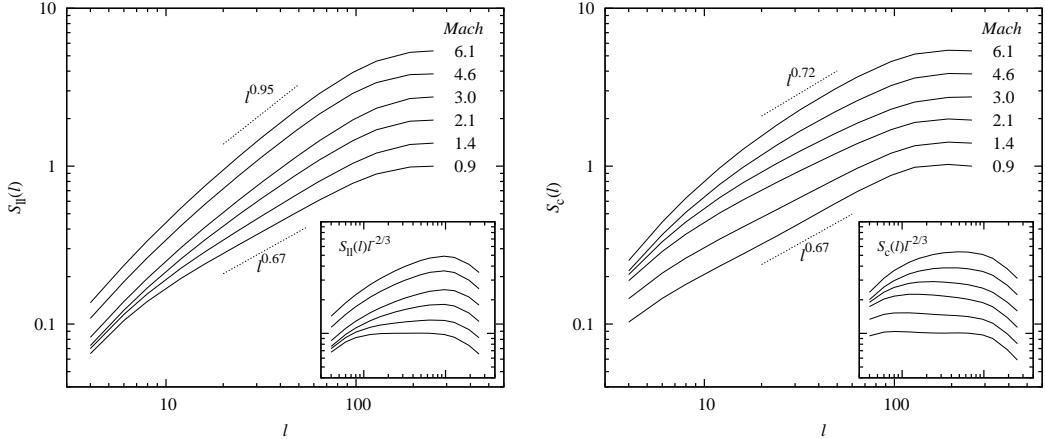}}
\caption{The longitudinal velocity structure function 
(left panel) and second order scalar structure function (right panel) 
at  different Mach numbers. The structure functions at $M=0.9$ 
are normalized to be unity at $l=256$, and, for clarity,  the curves 
are shifted upward for larger Mach numbers. The insets show 
the structure functions compensated by the $l^{2/3}$ scaling for 
incompressible turbulence, i.e., the Obukohov-Corrsin scaling 
for the scalar structure function and the Kolmogorov 
scaling for the velocity structure function.
}
\label{fig:SF}
\end{figure*}

\begin{figure*}
%\epsscale{1.}
%\plotone{f5.eps} 
\centerline{\includegraphics[height=6cm]{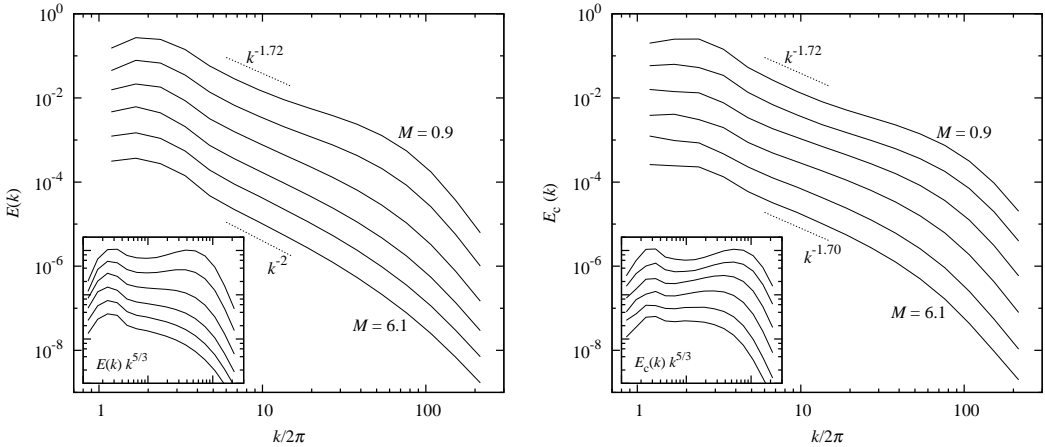}}
\caption{The scalar spectra (left panel) 
and the velocity spectra (right panel) at 
different Mach numbers. The spectra 
at $M=0.9$ are normalized so that the integration of the spectra 
is unity, and, for larger Mach numbers, the curves are shifted downward 
for clarity. The insets show the spectra  compensated by the $k^{-5/3}$ 
scaling for incompressible turbulence.}
\vspace{0.2in}
\label{fig:PS}
\end{figure*}

In Fig.\ \ref{fig:SF} we show the second order structure functions 
for the velocity and scalar fluctuations.
Note that these results are from volume-weighted averages,  
although using density weighting  would be more consistent 
with the dissipation timescales for the density-weighted 
velocity and scalar variances studied above. 
However,  unlike 1-point statistics, the choice of an 
appropriate density-weighting scheme is not trivial 
for 2-point statistical measures, and
will be pursued in future studies. Nevertheless, the 
volume-weighed structure functions and spectra shown 
here uncover interesting differences between the velocity and scalar 
structures.  

The curves in the left panel of Fig.\ \ref{fig:SF} are the longitudinal 
velocity structure functions, $S_{\rm ll} (l) \equiv \langle [(\bs{v}(\bs{x} +\bs{l})- \bs{v}(\bs{x})) \cdot \frac{\bs{l}}{l}]^2\rangle $,  
at different Mach numbers. The separation $l$ is in units of 
the resolution scale. At $M=0.9$,  an extended inertial 
range (about a decade) exists already at the resolution 
of 512$^3$. In this range, the structure function is well 
fit by a power-law with a slope of 0.67, consistent with the 
Kolmogorov scaling for incompressible turbulence. 

At larger Mach numbers, the structure function is
significantly steeper, as found in previous studies 
(e.g., Kritsuk et al.\ 2007). We find that the scale range 
with good power-law fits decreases with $M$, such that 
at $M= 6.1$ there is barely any range where the structure function 
is well fit by a straight line in the log plot. This decrease of 
the power-law range may be caused by stronger numerical 
viscosity in the Flash code to stabilize shocks at larger Mach 
numbers (see discussions below). To estimate the best-fit  
slopes of the structure functions at high Mach numbers, 
we choose a small range of $l$ from 20-70, far from both the dissipation range 
and the driving scale.  In this range, the exponents, $\zeta_{\rm v}$, 
are 0.67, 0.73, 0.83, 0.89, 0.92 and 0.95 for Mach numbers from 
0.9 to 0.61. 

The structure functions steepen with $M$ for two reasons. 
First, the number of shocks and the average shock intensity 
both increase with increasing Mach number, and shocks 
tend to give rise to a structure function that scales linearly 
with separation. This is why at large Mach numbers the scaling 
exponent approaches 1, which is sometimes referred to 
as the Burgers scaling (e.g., Kritsuk et al.\ 2007). 
Second, as discussed earlier, the pressure term preferentially 
converts kinetic energy into thermal energy.  Thus along the energy 
cascade some kinetic energy may be lost by the pdV work,  
resulting in smaller velocities toward small scales, and thus a 
steeper structure function. This effect may peak at $M \approx 2$ 
where the fraction of kinetic energy loss by pdV work is maximum.  
These two reasons are interrelated as pdV work converts kinetic 
energy to heat across shock  fronts.

The  2nd order structure function for the scalar field 
$S_{\rm c} (l) \equiv \langle (C(\bs{x} +\bs{l})- C(\bs{x}) )^2 \rangle$, 
is shown in the right panel of Fig.\ \ref{fig:SF}.  
Its behavior as a function of the Mach number is 
very different  from the velocity structure function. 
The scalar structure function first flattens as the Mach 
number increases from  0.9 to 2.  In this range of Mach numbers, 
we find an extended inertial-convective range for the scalar 
structure function, and the scaling exponents can be accurately 
measured.  The scaling exponent $\zeta_{\rm c}$ is 0.67 at $M=0.9$, 
in agreement with the Obukohov-Corrsin scaling,  
and it decreases to 0.62 and 0.59 at $M= 1.4$ and $2.1$, respectively. 

These values are in good agreement 
with the prediction from the cascade picture, 
represented by eq.\ (\ref{eq:exponents}). 
As found earlier,  the scaling exponents for the 
velocity structure function are $\zeta_{\rm v} = 0.73$ and $0.83$ at $M=1.4$ and $M=2.1$. 
Therefore, eq.\ (\ref{eq:exponents})  would predict 
$\zeta_{\rm c} = 0.64$ and 0.59 at these two Mach numbers, 
which are very close to the values measured from the simulation 
data. This confirms the validity of the cascade picture for 
scalars in flows with $M \lsim 2$. 
Physically, the flattening of the scalar structure function with $M$  
occurs because, as the  scaling of $\delta v(l)$ steepens at larger 
Mach number, there is relatively less velocity power at small scales, 
and thus the cascade of scalar fluctuations becomes slower, 
leading to a flatter scalar structure function and scalar spectrum. 

On the other hand, eq.\ (\ref{eq:exponents}) is no longer 
valid at Mach numbers larger than 2 where the 
scalar structure function starts to become steeper. The scaling exponents are $\zeta_{\rm c} =$ 
0.60, 0.67, and 0.72 for $M =3$, $4.6$, and 6.1, 
respectively.  As in the velocity case, the inertial-convective 
range becomes smaller at larger Mach numbers, 
and these exponents are measured in the range of $l \in$ [20-70]. More 
accurate measurement for these large Mach numbers 
requires higher resolution.
 
The reason for the steepening of the scalar structure 
functions is probably different from that for the 
velocity structure functions.  First, for scalars 
there is no counterpart for the effect of pdV work in 
the velocity case, which, as discussed above, 
can cause kinetic energy loss along the 
cascade and make the velocity scaling steeper. 
Instead, the scalar energy is exactly conserved along 
the scalar cascade. Second, although at large 
Mach numbers scalar ``edges"  can contribute to 
the steepening of the scalar structure function, it is not 
clear how much contribution these edges can make. 
Unlike the velocity shocks, they are not discontinuities 
(see \S 4.2),  and thus their contribution to steepening of the 
scalar structure functions may be smaller than that of 
shocks to that of the velocity structure functions.  
  
We think that the more important reason responsible 
for the steepening of the scalar structure function 
for $M \simeq 3$ is related to strong compressions and 
expansions in these flows. At large Mach numbers, 
strong compressions concentrate most of the mass 
in small post-shock regions, while most of the volume 
is occupied by expanding regions. In the expansion regions, 
the typical length scale of the scalar field increases as the scalar follows 
the expansion. On the other hand, the scalar length scale is 
reduced in compressed regions. Since the scalar structure function 
considered here is volume weighted, the effect of producing 
large structures by expansions dominates over the 
opposite effect by compressions. This leads to more 
power in  scalar fluctuations at large scales, 
and thus a steeper volume-weighted structure function. 
This tends to counteract the flattening of the scalar 
structure function discussed above, and starts to 
dominate at $M\gsim 3$ when the density 
fluctuations become very strong. We speculate 
that this effect, which may also contribute to the 
steepening of the volume-weighted velocity structure 
function, could be removed by 
introducing an appropriate density weighting scheme. 
However, while several previous studies have shown the 
importance of density-weighting in interpreting the statistics 
of supersonic turbulence (e.g.\ Kritsuk et al.\ 2007; Pan and Padoan 2009;  
Pan et al.\ 2009) it is still not clear what would be an 
appropriate density weighting factor for  two-point statistical measures
\footnotemark\footnotetext{For the moment, our speculation that density weighting could remove
the steepening at large Mach number is supported by 
preliminary calculations in which the  density-weighting 
factor is taken to be the average of the densities at the two points. 
With this weighting, we found that the 2nd order scalar structure 
functions do not significantly steepen at $M\gsim 3$. However, 
applying the same weighting to the velocity structure 
functions seems to give strange results. 
To find an appropriate density weighting factor for both 
the velocity field and the scalar field,
further investigation is needed, which should perhaps aim 
at determining whether a scheme exists in which eq.\ (\ref{eq:exponents}) is valid for all 
Mach numbers.}.  The topic of density-weighting for 
structure functions is out of the scope of the 
current paper and will be explored separately in a 
future study.

To summarize  these results, we find that for $M \lsim 2$ the 
scaling exponents  for the scalar and velocity structure 
functions agree well with eq.\ (\ref{eq:exponents}), supporting the cascade picture for
 the scalar fluctuations at these Mach numbers. 
On the other hand, the steepening of the scalar structure function 
with the Mach number for $M > 2$ is in contradiction to 
the prediction by eq.\ (\ref{eq:exponents}). 
However, this does not mean that the cascade picture 
is invalid for scalars in high Mach number flows, but 
rather suggests the need for an appropriate density-weighting scheme 
when studying their statistics.
In fact,  the validity of the cascade picture at all Mach numbers 
is supported by our results for the mixing timescale and its 
scale dependence.

Fig.\ \ref{fig:PS} shows the power spectra of both the 
velocity and scalar fields, again employing volume 
weighted averages. At the current resolution, there 
is barely any inertial range in the spectra for all Mach numbers. 
At small Mach numbers, there exist strong bumps, named 
bottlenecks (Falkovich 1994; Kritsuk 2007), close to the dissipation 
range (see the compensated spectra in the insets). The bottlenecks 
appear to be weaker at larger Mach numbers.  
This suggests that the numerical scheme in 
the FLASH code probably applies a larger numerical viscosity to stabilize shocks 
at larger Mach numbers.  We find that essentially no bottlenecks 
exist at $M \gsim 3$, which is different from simulations 
with less diffusive numerical codes (e.g., Kritsuk et al. 2007; 
Lemaster and Stone 2009). As mentioned above, strong 
numerical diffusion at high Mach numbers may be 
responsible for the reduction in the inertial scale range 
with good power-law fits for the structure functions 
shown in Fig.\ \ref{fig:SF}. However, as discussed 
below, we find evidence that strong numerical 
diffusion in the FLASH code at large $M$ does 
not affect our conclusions on the behavior 
of the power spectra and structure functions 
in the inertial range.    

Due to the very limited inertial range in the spectra 
in Fig.\ \ref{fig:PS} at all Mach numbers, 
an accurate measurement of the spectra slope 
is not possible. However, the overall trend in the 
power spectra as a function of Mach number is 
the same as that in the structure functions. 
As can be seen from the insets, the velocity spectra 
keeps steepening with increasing $M$, while the 
scalar spectra flattens at first and starts to be 
steeper at $M \gsim 3$. Also similar to the 
structure functions, the scalar spectrum 
is significantly flatter than the velocity 
spectrum at all Mach numbers except  the $M=0.9$ 
case where the slopes are about equal. 
Future higher-resolution simulations are 
needed to quantify these results more 
accurately.

Finally, we discuss the potential effects 
of  strong numerical diffusion in the 
FLASH code at large Mach numbers.  
We will show that our conclusions 
above are actually not affected. 
Due to the limited inertial range at  
current resolution, strong numerical diffusion 
may give rise to overestimates in the slopes 
of the power spectra and structure functions. 
Consider the structure functions for example. 
A large numerical viscosity  
would reduce the inertial range, and 
increase the chance that some of the 
scales in the range we chose to measure 
the structure functions 
would be affected by numerical diffusion. 
This may overestimate the slope of 
the structure function. Therefore, one may 
suspect that the steepening of the structure 
function in the inertial range found in our 
simulations is due to larger numerical 
viscosity at larger Mach numbers. 
We find that is not case by comparing 
our results at $M=6$ to those from simulations by 
Kristuk et al. (2007) using less diffusive codes, who find 
prominent bottlenecks in their velocity spectra.

Kritsuk et al. (2007) found a slope of 0.95 for 
the volume-weighted longitudinal structure 
function in the inertial range of a $M=6$ 
flow. This slope is exactly the same as 
that measured in the scale range of [20-70] 
resolution scales in our simulations at the 
same $M$. This suggests that the scale 
range we chose is essentially independent 
of bottlenecks. As seen in Fig.\  7 of Kritsuk et al. (2007), 
bottlenecks may contaminate the velocity 
spectra only below $\sim$ 30 resolution scales.  
At those small scales, i.e., below 20-30 times the resolution scale, 
the velocity structure function in our $M=6$ flow is 
steeper than that in Kristuk et al. (2007), consistent 
with stronger numerical diffusion 
in the FLASH code at large $M$.  
In conclusion, above $\sim 20$ resolution scales, 
the slopes of the structure functions 
are not affected by bottlenecks, and the 
steepening trend at large $M$ found in our 
simulations is realistic. This conclusion also 
applies to the scalar structure functions 
since the numerical diffusion for scalars in the 
FLASH code is essentially the same as the numerical viscosity.
  
Furthermore, a comparison of the velocity 
spectrum in our $M=6$ flow to that in Kritusk et al. 
(2007) shows that the stronger numerical 
viscosity in our simulations makes the 
spectrum slightly steeper (by a few percent).
However, this uncertainty of a few percent 
is smaller that the slope change from $M=3$ to 
$M=6.1$ found in our simulations, suggesting 
that the steepening of the velocity spectrum 
at large $M$ is also realistic.  

\subsection{Scalar PDF}

\begin{figure*}
%\epsscale{1.}
%\plotone{f4.eps} 
\centerline{\includegraphics[height=6cm]{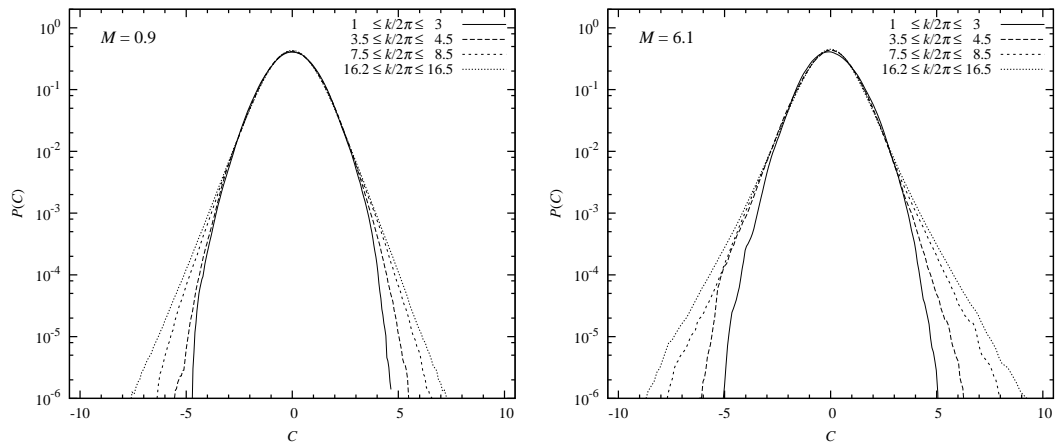}}
\caption{The concentration PDFs for scalars forced at different length scales.   
The left panel and the right panel are for $M=0.9$ and $M=6.1$, 
respectively. The PDFs are normalized to have unit variance.  Different 
curves correspond to scalars forced at different ranges of wave numbers.}
\vspace{0.2in}
\label{fig:PDF}
\end{figure*}

In this subsection, we study the probability distribution function 
(PDF) of forced scalars in compressible turbulence. Here we first 
calculated the 1-point density-weighted PDF, $P(C, t)$, 
in each snapshot by $P(C, t) = \frac{1}{V} \int_V \delta(C-C(\bs{x},t))\tilde\rho(\bs{x},t) d\bs{x}$ 
where $V$ the volume of the simulation box. 
Then for each Mach number we obtained the average 
PDF, $P(C)$, over 10 snapshots covering about 10 
dynamical times. The results for Mach numbers 0.9 and 6.1 are shown Fig.\ \ref{fig:PDF}, 
where all the PDFs are normalized to have unity variance.
Note that the PDFs shown here are not  the PDFs of the 
scalar differences between two points at different separations, 
which will not be considered in the present paper.

The four lines in each panel of this figure correspond to  
different forcing schemes for the scalars.  We find that,  
although the scalar source term is set to be Gaussian 
in all our simulations, the scalar PDF is Gaussian only 
if the scalar is forced at the flow driving scale. This agrees 
with the Gaussian scalar PDF found in Watanabe \& Gotoh (2004) 
for incompressible turbulence where the scalar source 
was also injected at the flow driving scale. 
For scalars forced at smaller scales, the PDFs show 
broad non-Gaussian tails, which  become broader 
with decreasing source length scale at all Mach 
numbers.

This behavior is likely to be related to the PDF 
of the velocity difference as a function of 
separation. The velocity PDF at the driving scale 
is nearly Gaussian, and the PDF for the scalar 
forced at this scale is nearly Gaussian as well.
On the other hand, the PDF of the velocity 
difference across a small separation is 
known to have broad tails, which become 
broader as the separation decreases, a phenomenon 
known as intermittency (e.g., Frisch 1995). Since the 
scalar cascade is controlled by the velocity 
field, such small-scale non-Gaussian 
velocity structures are likely to leave a signature on the 
PDF tails of scalars forced at small scales. Furthermore the 
small-scale velocity structures are known to become more 
intermittent at larger Mach numbers (see,  e.g., Padoan et al.\ 2004). 
Again, this may be directly responsible for the fact that the PDF 
tails for scalars forced at small scales broaden with 
increasing Mach number, as can be seen by comparing
the two panels in Fig.\  \ref{fig:PDF}. 

We point out that, although the non-Gaussian 
velocity structures contribute to the broad tails in 
the 1-point PDFs of scalars forced at small scales,  
it is not clear whether the non-Gaussianity in the 
velocity structures is the only cause for non-Gaussian 
tails in the scalar PDFs. 
In other words, it is possible that the scalar PDFs 
would exhibit non-Gaussian tails even if the flow 
velocity were Gaussian at all scales, and that 
non-Gaussian velocity structures only enhance an 
already existing effect. In fact, previous studies of 
mixing in incompressible turbulence 
show that at small scales the scalar difference  PDF is 
non-Gaussian even if the advecting velocity field is exactly 
Gaussian (see, e.g., Shraiman \& Siggia 2000), 
meaning that the non-Gaussianity in the scalar 
difference statistics is an intrinsic feature of 
mixing itself. Whether this is also true for 
the 1-point PDFs of scalars forced at small 
scales is a question for future studies.  

\subsection{Decaying Scalars}

\begin{figure*}
%\epsscale{1.}
%\plotone{f4.eps} 
\centerline{\includegraphics[height=6cm]{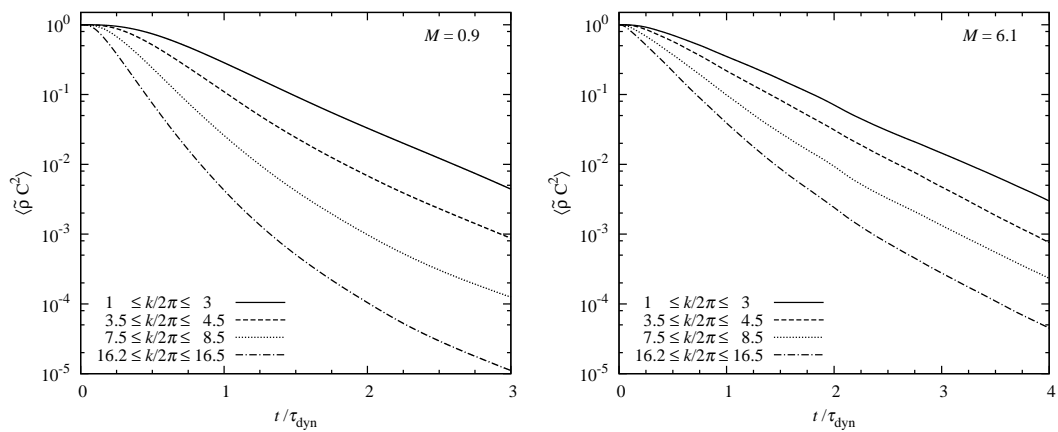}}
\caption{The concentration variance as a function of time for decaying 
scalars in turbulent flows at $M=0.9$ (left panel) and $M=6.1$. Different 
curves in each panel correspond to different length scales (or initial spectra) 
of the initial scalar field.}
\vspace{0.1in}
\label{fig:decay} 
\end{figure*}

Finally, we consider the evolution of decaying scalar 
fields with no continuous sources (i.e., pollutants 
added to the flow only once). As mentioned earlier, 
we included four decaying scalars in our simulations, each of which has a 
different initial spectrum (or length scale). Fig.\ \ref{fig:decay} 
shows the decay of the density-weighted concentration 
variance for the four scalars, focusing on the 
results for $M=0.9$ and $M=6.1$. The results for 
other Mach numbers are qualitatively similar.  
The variance decay as a function of time 
and its dependence on the initial scalar 
scale is similar to that found by Eswaran 
and Pope (1988) for the incompressible case (see their Fig.\ 10). 
Each curve in Fig.\ \ref{fig:decay} has a 
short initial period where it is flat and smooth. 
During this period, the cascade develops 
scalar structures toward smaller scales, and true 
mixing by molecular diffusion (numerical diffusion in 
our simulations) simply waits for the 
structures to reach the diffusion scale.   
Once structures of the size of the diffusion 
scale appear,  the variance starts to decay steadily.

We find that the concentration variance can be 
approximately fit by a single exponential 
function for decaying scalars with initial length 
scale equal to the  flow  driving scale 
(see $1 \le k/2\pi \le 3$ curves in Fig.\ \ref{fig:decay}). 
This is true for any Mach number.  
Fitting the variance decay of these scalars by 
exponentials shows that the decaying timescale
is in the range of $0.5-0.6 \tau_{\rm dyn}$ for 
$M \approx 1$ to 6. The change of the decay 
timescale with $M$ is similar to that for the 
forced cases, i.e., it increases with $M$ 
at small $M$ and becomes constant at 
$M \gsim 3$. We note that these timescales 
are smaller than those for the corresponding forced cases 
(see Table 2).  The reason for this is that, without continuous sources 
at large scales, the spectrum of a decaying scalar is 
flatter than that of a forced scalar, and, with relatively 
more fluctuations at smaller scales, the mixing process is faster.  

For scalars with initial scales smaller than 
the flow driving scale, the variance 
evolution cannot be fit by a single exponential function 
(with the $3.5 \le k/2\pi \le 4.5$ curve for $M=6$ 
being the only exception). The decay generally 
consists of two phases, an early fast phase 
and a slower phase that starts at a few 
dynamical timescales.   
In the fast decay phase, the timescale 
for the variance decay decreases with decreasing 
initial scalar scale, consistent with the scale 
dependence of the mixing timescale in the 
forced case. For example, roughly fitting 
exponentials to the fast phases of the 
bottom three curves in the $M=6.1$ panel of 
Fig.\ \ref{fig:decay}, we find that the timescales,  from top to bottom, are respectively, 
$0.56 \tau_{\rm dyn}$, $0.4 \tau_{\rm dyn}$ and $0.3 \tau_{\rm dyn}$. 
Again, these timescales for the variance decay are  
smaller than the corresponding timescales 
for the forced cases. In the fast phase, 
the scalar fluctuations are significantly homogenized, 
with the variance reduced by a factor of 100 -1000 
in a few dynamical timescales.

\begin{figure*}
%\epsscale{1.}
%\plotone{f4.eps} 
\centerline{\includegraphics[height=6cm]{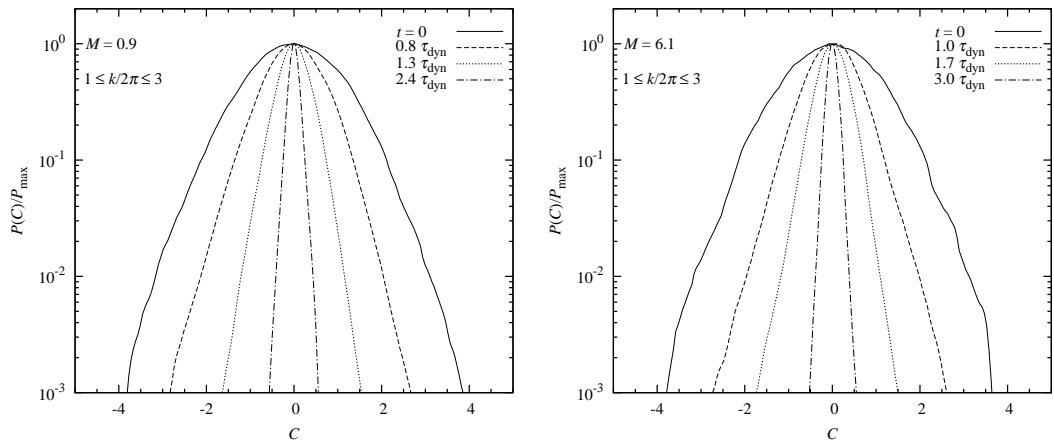}}
\caption{The density-weighted concentration PDF as a 
function of time for a decaying scalar in turbulent flows at $M=0.9$ (left panel) 
and $M=6.1$. The scalar field has an initial spectrum in the 
$1 \le k/ 2\pi \le 3$ range. For both cases, the initial pdf is 
approximately Gaussian with unity variance.  All the pdfs 
are normalized such that their maxima are unity.}
\vspace{0.1in}
\label{fig:decaypdf} 
\end{figure*}

After that, a slower phase starts, which corresponds to the 
evolution of the scalar spectrum. We find that if the initial 
scalar spectrum is within the inertial range of the flow, 
the spectrum spreads out in {\em both} directions, i.e., 
toward smaller scales, due to the cascade, 
and also  toward larger scales. The spread to larger scales is 
due to the fact that interactions of scalar fluctuations at 
a given scale with velocity fluctuations at larger 
scales can produce scalar fluctuations at larger 
scales.  We refer to this effect as the backward transfer.  
In a few dynamical timescales, the scalar power at 
small scales is significantly erased by mixing, and the 
fluctuation power starts to be dominated by structures 
at larger scales caused by the backward transfer. This 
results in an increase in the characteristic length scale 
for the scalar fluctuations, leading to slower mixing rate at later times.  
The backward transfer is expected to stop when 
the typical scalar length scale becomes close to 
the driving scale of the flow, $L_f$, beyond which 
there are no considerable velocity power. 
Therefore, in the later phase, the scalar length scale is 
always close to the flow driving scale, and the decay 
time is approximately given by the turnover 
time of the eddies at the flow driving scale. 
This is why at late time the decay rates for 
all scalars tend to become similar.

This picture also explains why the variance of a 
decaying scalar can be approximately fit by a 
single exponential function if its initial length 
scale is close to the flow forcing scale. In that case,  
the characteristic length scale of the scalar is 
fixed around that flow driving scale and does 
not change  with time.   

In summary, the timescale for the variance decay 
is smaller by 30-40 \% than that obtained in the 
corresponding forced case. It is about 0.5-0.6 
dynamical timescale if the initial scalar length 
scale is close to the flow driving scale. The dependence 
of the decay timescale on the initial length scale and 
on the Mach number for decaying scalars is 
generally consistent with that in the forced case.

Fig.\ \ref{fig:decaypdf} shows the evolution of the density 
weighted scalar probability distribution, $P(C,t)$, for the 
scalar whose initial spectrum is in the range 
$1 \le k/ 2\pi \le 3$, i.e., the same as the flow forcing 
spectrum.  The initial PDFs are approximately Gaussian. 
The PDFs become narrower with time, 
indicating homogenization of the scalar fluctuations. 
We point out that Fig.\ \ref{fig:decaypdf} 
is only meant to illustrate the general problem of 
how the PDF of a decaying scalar evolves. In practical 
applications, the initial PDF is typically not Gaussian, 
but instead is likely to be bimodal with two peaks, 
representing the unmixed pollutants and unpolluted 
flow, respectively (see, e.g., Pan 2008).

In Fig.\ 8, we see that at later times exponential tails 
develop in the PDFs. Note that in the corresponding 
forced cases, i.e., for scalars forced at the flow driving 
scale, the concentration PDF shown in Fig.\ 6  is 
Gaussian (see discussion below). A likely origin for the 
fat  tails in the PDFs of decaying scalars shown in Fig. 8 
at the later time is that mixing of concentration levels 
at the PDF tails is more difficult than that for the central 
part. Significantly moving the tails toward the central part 
requires contact and mixing between fluid elements 
with extreme concentrations corresponding to opposite 
tails of the scalar PDF, i.e., one from the high tail and 
the other from the low tail. Clearly, these events are 
much rarer than those for the mixing of the central part 
of the PDF. This suggests that the tails are more persistent, and this
persistency tends to give fat tails.   

Exponential tails also exist in the PDFs of 
decaying scalars with smaller initial length 
scales. Comparing the PDFs of different 
decaying scalars, we find that, at similar 
variances (which correspond to different 
evolution times for different scalars 
since their decaying timescales are different),  
the PDF tails are broader for scalars with 
smaller initial length scales. In other words, 
the exponential tails develop faster 
for decaying scalars with smaller 
initial length scales. The breadth of 
the tails also increases with increasing 
Mach numbers. These trends are consistent 
with those found for forced scalars. Similarly 
this trend is related to the non-Gaussian 
velocity structures at small scales. 

The exponential tails for the decaying 
scalars with initial scale at the flow 
driving scale raises the question why the PDFs 
of scalars continuously forced at the flow driving scale 
is Gaussian, considering that scalar sources injected 
at early times have ``decayed" and would 
thus give exponential contributions to the 
PDFs at the current time.      
The answer relies on the fact that the main 
contribution to the PDF of a forced scalar 
at a given time is from sources injected 
within about a mixing timescale since 
the older sources have already been 
significantly mixed. We find that a decaying 
scalar with initial scale at the flow driving 
scale develops considerable exponential 
tails only after more than a mixing timescale.    
This explains why the PDFs in the 
corresponding forced cases are Gaussian. 
On the other hand, for scalars forced at 
smaller scales, recent scalar sources 
have larger contributions to the tails of 
the PDFs at the current time because 
fat tails develop faster for sources 
injected at smaller scales. This is 
consistent with exponential tails for scalars 
forced at small scales as found in Fig.\ 6.

\section{Discussion: the Effect of Compressible Modes}

Our calculations for the energy dissipation 
timescale and the mixing timescale in \S 4.3 
have shown that
the dissipation of kinetic energy becomes faster 
at higher Mach numbers, while the dissipation of 
scalar fluctuations becomes slower if there is more 
kinetic energy contained in compressible modes. 
This result has two interesting implications. First, the 
compressible modes significantly enhance energy 
dissipation, but not scalar dissipation. Second, 
the compressible modes are slower at 
enhancing mixing than the solenoidal modes. 
We give physical reasons for these two points.  

The key to the first point is the different 
effects of shocks on energy dissipation
and scalar dissipation. As mentioned in \S4, every 
shock is a strongly dissipative structure for
kinetic energy,  
and thus the existence of shocks  significantly contributes 
to energy dissipation. On the other hand, the effect 
of shocks on mixing is much weaker. 
Shocks do not give rise to scalar discontinuities, 
instead they amplify scalar gradients by a finite 
factor which is equal to the density jump. For a shock to 
produce a scalar structure that can dissipate the 
scalar fluctuations at the same level as it 
dissipates energy, it has to be strong 
enough to reduce the scalar length scale 
to around the diffusion scale, 
so that molecular diffusion can efficiently 
mix right after the compression. Since the diffusion 
scale is typically much smaller than the scalar injection 
scale, only the strongest shocks or successive 
compressions by strong shocks from different 
directions could possibly produce such structures.      

We find that such strong shocks are very 
rare. Consider our simulations for example. 
For the scalar forced at the flow driving 
scale, a compression by a factor of $\sim$100 
is needed to bring a scalar structure at the 
source scale down to the diffusion scale, 
since the diffusion scale in our simulations 
is around the resolution scale, and the 
resolution is 512$^3$. The frequency of 
such strong shocks may be characterized 
by the fraction of regions with density 100 times 
larger than the average because the scalar 
scale reduction by shocks follows the density jump. 
Using the density PDF in our simulated flow 
at Mach 6,  the mass fraction of these 
regions is $\sim 10^{-4}$. This is much 
smaller than the mass fraction of regions 
compressed by intermediate 
or weak shocks. For example,  
regions compressed by shocks 
with a jump factor larger than 10 
make up a mass fraction of about 0.1,  3 
orders of magnitude larger than that of regions 
affected by shocks with a jump factor of 100. 
Therefore, strong shocks that can 
directly produce structures at the 
diffusion scale are rare. The probability is even smaller 
at lower Mach numbers. In realistic environments 
with much larger Reynolds and Peclet 
numbers, the diffusion scale relative to 
the source scale is much smaller, while the density jump is probably 
independent of the viscosity or the Reynolds number. 
Therefore at realistic Reynolds and Peclet numbers, 
there would be essentially no shocks that are strong 
enough to reduce the scalar structure directly to 
the diffusion scale. We thus conclude that shocks 
generally do not produce scalar structures that 
dissipate scalar fluctuations as strongly as 
they dissipate kinetic energy. This is responsible for 
the different roles of  compressible modes 
in energy dissipation and scalar dissipation.

We next show that the compressible modes are 
less efficient at enhancing mixing than the solenoidal 
modes. As mentioned earlier, compressible modes 
include both compressions and expansions, and 
compressions amplify scalar gradients, while 
expansions reduce the scalar gradients. 
Because the impact of compressions and 
expansions on the scalar gradient is proportional
to the density change, the overall effect of 
compressible modes on the density-weighted gradient 
variance, $\langle \tilde{\rho} \partial_i C \partial_i C \rangle$, 
may be estimated 
as $\langle \rho^3 \rangle/\bar{\rho}^3$ (where one factor of 
$\rho/\bar{\rho}$ is from density-weighting). We find that this factor 
is about 90 in our flow at $M=6.1$.  This suggests that, 
along with the development of the density fluctuations, 
which takes about a dynamical time, the variance of 
the scalar gradients may be amplified by a factor 
of 90 by compressible modes in about a dynamical timescale. 

Although a factor of 90 seems significant, it is small
compared to the enhancement of  the scalar gradients 
achieved by solenoidal modes. Based on the 
mixing timescale in incompressible flows, 
incompressible modes can reduce the 
scalar structures to the diffusive scale in about a 
dynamical timescale. Consider our simulations 
again as an example. The diffusion scale in our 
simulations is roughly 1/100 of the source length scale, 
and thus the incompressible turbulent flow can increase the 
scale gradient variance by a factor of $10^4$ in about a 
dynamical timescale. This is 100 times more efficient than 
compressible modes in the $M=6.1$ flow. 
In other words, the contribution from compressible 
modes to the total scalar gradient amplification is at 
the level of $\sim$ 1\% in our simulations for $M=6.1$.
It is even smaller at smaller $M$ where the 
amplitude of density fluctuations is smaller. 
As pointed out earlier, this may be responsible 
for the slight decrease of the normalized 
mixing timescale in the range $3\le M \le 6$ shown in Fig.\ 2. 

The contrast between the contributions from the two 
modes will be even stronger with increasing Reynolds and Peclet 
numbers. At larger Peclet numbers, the solenoidal 
modes would reduce the scalar scale to smaller sizes 
in a dynamical timescale, giving a much larger amplification in the scalar gradients. 
On the other hand, the gradient enhancement by compressible 
modes would probably remain the same since the width 
of the density PDF does not increase with Reynolds number,
and is likely to have already converged at the 
resolution of 512$^3$ (e.g., Lemaster \& Stone 2008). The more 
efficient generation of small structures by solenoidal modes is probably 
because stretching by shears and vortices can continuously 
reduce the scalar structures to progressively small scales.  
On the other hand, the degree of compression is limited 
by gas pressure, which prevents the same fluid element 
from being compressed continuously, and the existence of 
expansions also tends to counteract the effect of compressions.  
This suggests that the solenoidal modes are more 
efficient at amplifying the scalar gradients than the 
compressible modes, which is responsible for 
the decrease of the mixing efficiency when 
more kinetic energy is contained in compressible modes. 
  
\section{Conclusions}

Mixing in compressible turbulence plays a key role 
in determining, for example, the metallicity dispersion 
of star forming regions, the abundance scatter in 
coeval field stars, and the transition from primordial 
to PopII star formation. Yet, there have been surprisingly few
theoretical or numerical studies of this process.

We have conducted the first systematic numerical 
study of the physics of mixing in supersonic 
turbulence. Here we give a summary of main 
results of our study.     

\begin{enumerate}

\item
If the typical length scale of the scalar 
source is close to the flow driving scale, 
the mixing timescale is similar to the 
timescale for the kinetic energy dissipation at all Mach 
numbers. Furthermore, both of these timescales 
are on the order of the dynamical timescale, which 
suggests that the generation of small-scale scalar fluctuations is 
through a cascade similar to that of the kinetic energy.

\item
The fraction of kinetic energy contained in compressible 
modes increases with Mach number for $M \lsim 3$, 
and becomes constant at larger $M$. For kinetic energy 
in the inertial range, the fraction is 1/3 for $M \gsim 3$, 
indicating an equipartition between compressible modes 
and solenoidal modes. Equipartition is established 
at each wave number in the inertial range 
of flows with $M \gsim 3$.

\item
The mixing timescale normalized to the dynamical 
timescale increases with increasing Mach number 
for $M \lsim 3$, and is essentially constant at 
larger Mach numbers where the fraction of kinetic
energy contained in compressible modes is constant. 
Compressible modes are less efficient 
at enhancing mixing, because they have 
a much lower efficiency at amplifying scalar 
gradients and producing small-scale scalar structures than the solenoidal modes, 
which are the primary ``mixer"  at all Mach numbers.  
Solenoidal modes contain most of kinetic energy even 
at very high Mach numbers, and the difference in the 
normalized mixing timescale at all Mach numbers 
is smaller than $\approx 20\%$. 

\item
As the characteristic length scale of the 
scalar sources decreases, the mixing timescale also 
decreases, and the dependence of the timescale 
on the source length scale is determined by 
the turnover time of eddies at the source scale. 
The dependence of mixing timescale on the 
source scale is weaker at larger Mach 
numbers, because the velocity scaling is steeper in 
these cases, and the decrease in the 
eddy turnover time with the eddy size is weaker.

\item
The timescale for the viscous dissipation of the kinetic 
energy, normalized to the dynamical timescale, decreases 
with the Mach number.  This is because compressible 
modes provide an additional fast channel for the 
dissipation of kinetic energy.  

\item
Significant kinetic energy loss in supersonic isothermal 
turbulence is contributed by pdV work for Mach number 
in the range $0.9 \lsim M \lsim 6.1$. The fraction of energy 
loss by pdV work peaks at $M \approx 2$. 

\item
The 2nd order velocity structure function steepens 
from the Kolmogorov scaling (i.e., a $2/3$ power law) 
to the Burgers scaling (i.e., a linear scaling) as the Mach 
number increases from subsonic to highly supersonic. 

\item
The 2nd order scalar structure function obeys 
a 2/3 power law in subsonic flows, in agreement 
with the Obukohov-Corrsin law for mixing in 
incompressible turbulence.  As the Mach number 
increases, the scalar structure 
function first flattens, and the scaling exponent decreases 
to about 0.6 at $M\simeq 2$. The measured 
scaling exponents in this range of $M$ are consistent 
with the prediction from the cascade picture with the 
measured exponents for the velocity structure functions.  
As the Mach number increases further, the scalar structure 
function becomes steeper, and the relation between the
scalar and the velocity scalings from the cascade 
picture is not obeyed.  However, this probably does not 
mean the scalar cascade picture is not valid; instead it 
motivates a future study to find an appropriate density-weighting scheme to compensate 
the effect of compressions and expansions 
on the scalar and velocity structure functions.     

\item 
The PDFs of the forced scalars are generally non-Gaussian 
even if the scalar sources are set to be Gaussian in 
our simulations. The scalar PDF is found to be 
Gaussian only for the scalar forced at the same length 
scale as the flow driving scale. The PDFs show broad tails 
for scalars forced at smaller scales, and the tail broadens 
with decreasing source length scale. 
The tails are also broader at larger Mach numbers.    

\item 
For decaying scalars,  the timescale for the scalar variance 
decay is slightly faster than the dissipation timescale  in 
the corresponding forced case. The decay timescale for a scalar with 
an initial  length scale close to the flow driving scale  is in the range 
$0.5- 0.6 \tau_{\rm dyn}$. The dependence of the variance decay 
timescale on the initial scalar length scale and on the 
Mach number is similar to that in the force cases. 
The PDF of  a decaying scalar narrows with time and 
displays non-Gaussian tails.
\end{enumerate}

While these conclusions shed insight on mixing in astrophysical 
environments, they also bring up new questions and point the 
way for further investigations. In this study, we have analyzed 
only low-order statistics for the velocity and the 
scalar structures in simulations with a moderate resolution  
of $512^3$. Examination of higher order statistics for these 
structures require higher resolution simulations, which will help 
probe deeper into the important physics of mixing in supersonic 
turbulence.
    
\acknowledgements

We are grateful to Robert Fisher for his helpful discussions. We thank the 
referee, Paolo Padoan, for comments and suggestions to improve the paper.   
We acknowledge the support from NASA theory grant NNX09AD106.
All simulations were conducted on the ``Saguaro'' cluster operated by the
Fulton School of Engineering at Arizona State University.
The results presented here were produced using the FLASH code, a 
product of the DOE ASC/Alliances-funded Center for Astrophysical Thermonuclear Flashes at the
University of Chicago.

\end{document}